\documentclass[twocolumn]{aastex631}

\usepackage{multirow}

\begin{document}

\title{Investigating the Nested Structure of the Outflow from the Low Luminosity Protostar IRAS 16253-2429 using JWST and ALMA}

\correspondingauthor{Mayank Narang}
\email{mayankn1154@gmail.com}
\author[0000-0002-0554-1151]{Mayank Narang\textsuperscript{\dag}}
\affiliation{Academia Sinica Institute of Astronomy and Astrophysics, 
11F Astronomy-Mathematics Building, AS/NTU, 
No.\ 1, Sec.\ 4, Roosevelt Rd, Taipei 106216, Taiwan}

\altaffiliation{\dag\ Currently postdoctoral fellow at Jet Propulsion Laboratory, California Institute of Technology}

\author[0000-0002-9497-8856]{Himanshu Tyagi}
\affiliation{Department of Astronomy and Astrophysics Tata Institute of Fundamental Research \\Homi Bhabha Road, Colaba, Mumbai 400005, India}

\author[0000-0003-0998-5064]{Nagayoshi Ohashi}
\affiliation{Academia Sinica Institute of Astronomy and Astrophysics, 11F of Astronomy-Mathematics Building, AS/NTU, \\ No.\ 1, Sec.\ 4, Roosevelt Rd, Taipei 106216, Taiwan}

\author[0000-0002-3530-304X]{P. Manoj}
\affiliation{Department of Astronomy and Astrophysics Tata Institute of Fundamental Research \\Homi Bhabha Road, Colaba, Mumbai 400005, India}

\author[0000-0001-7629-3573]{S. Thomas Megeath}
\affiliation{Ritter Astrophysical Research Center, Department of Physics and Astronomy, \\ University of Toledo, Toledo, OH 43606, USA}

\author[0000-0002-6195-0152]{John J. Tobin}
\affil{National Radio Astronomy Observatory, 520 Edgemont Rd., Charlottesville, VA 22903 USA}

\author[0000-0001-7591-1907]{Ewine F. Van Dishoeck}
\affiliation{Leiden Observatory, Universiteit Leiden, Leiden, Zuid-Holland, NL}
\affiliation{Max-Planck Institut f\"ur Extraterrestrische Physik, Garching bei München, DE}

\author[0000-0001-5175-1777]{Neal J. Evans II}
\affiliation{Department of Astronomy, The University of Texas at Austin, \\ 2515 Speedway, Stop C1400, Austin, Texas 78712-1205, USA}

\author[0000-0001-8302-0530]{Dan M. Watson}
\affiliation{University of Rochester, Rochester, NY, US}

\author[0000-0001-8876-6614]{Alessio Caratti o Garatti}
\affiliation{INAF-Osservatorio Astronomico di Capodimonte, IT}

\author[0000-0001-9133-8047]{Jes K. J\o rgensen}
\affiliation{Niels Bohr Institute, University of Copenhagen, \O ster Voldgade 5-7, 1350, Copenhagen, Denmark}

\author[0000-0002-6447-899X]{Robert Gutermuth}
\affiliation{University of Massachusetts Amherst, Amherst, MA, US}

\author[0000-0002-8238-7709]{Yusuke Aso}
\affiliation{Korea Astronomy and Space Science Institute, 776 Daedeok-daero,  Yuseong-gu, Daejeon 34055, Republic of Korea}
\affiliation{Division of Astronomy and Space Science, University of Science and Technology, \\ 217 Gajeong-ro, Yuseong-gu, Daejeon 34113, Republic of Korea}

\author[0000-0002-1700-090X]{Henrik Beuther}
\affiliation{Max Planck Institute for Astronomy, Heidelberg, Baden Wuerttemberg, DE}

\author[0000-0002-4540-6587]{Leslie W. Looney}
\affiliation{Department of Astronomy, University of Illinois, 1002 West Green St, Urbana, IL 61801, USA}
\affil{National Radio Astronomy Observatory, 520 Edgemont Rd., Charlottesville, VA 22903 USA} 

\author[0000-0001-8341-1646]{David A. Neufeld}
\affiliation{William H. Miller III Department of Physics and Astronomy, The Johns Hopkins University, Baltimore, MD, USA}

\author[0000-0002-7506-5429]{Guillem Anglada}
\affiliation{Instituto de Astrof{\'i}sica de Andaluc{\'i}a, CSIC, Glorieta de la
Astronom{\'i}a s/n, E-18008 Granada, ES}

\author[0000-0002-6737-5267]{Mayra Osorio}
\affiliation{Instituto de Astrof{\'i}sica de Andaluc{\'i}a, CSIC, Glorieta de la Astronom{\'i}a s/n, E-18008 Granada, ES}

\author[0000-0001-8790-9484]{Adam E. Rubinstein}
\affiliation{University of Rochester, Rochester, NY, US}

\author[0000-0002-6136-5578]{Samuel Federman}
\affiliation{Ritter Astrophysical Research Center, Department of Physics and Astronomy, \\ University of Toledo, Toledo, OH 43606, USA}
\affiliation{INAF-Osservatorio Astronomico di Capodimonte, IT}

\author[0000-0003-1430-8519]{Lee W. Hartmann}
\affiliation{Department of Astronomy, University of Michigan -- Ann Arbor, Ann Arbor, MI 48109, USA}

\author[0000-0002-4448-3871]{Pooneh Nazari}
\affiliation{Leiden Observatory, Universiteit Leiden, Leiden, Zuid-Holland, NL}

\author[0000-0003-3682-854X]{Nicole Karnath}
\affiliation{Space Science Institute, Boulder, CO, US}
\affiliation{Center for Astrophysics Harvard \& Smithsonian, Cambridge, MA, US}

\author[0000-0002-8115-8437]{Hendrik Linz}
\affiliation{Max Planck Institute for Astronomy, Heidelberg, Baden Wuerttemberg, DE}
\affiliation{Friedrich-Schiller-Universit\"at, Jena, Th\"uringen, DE}

\author[0000-0002-5812-9232]{Thomas Stanke}
\affiliation{Max-Planck Institut f\"ur Extraterrestrische Physik, Garching bei München, DE}

\author[0000-0001-7491-0048]{Tyler L. Bourke}
\affiliation{SKA Observatory, Jodrell Bank, Lower Withington, Macclesfield SK11 9FT, UK}

\author[0000-0001-8227-2816]{Yao-Lun Yang}
\affiliation{RIKEN Cluster for Pioneering Research, Wako-shi, Saitama, 351-0106, Japan}

\author{Rolf Kuiper}
\affiliation{Faculty of Physics, University of Duisburg-Essen, Lotharstra{\ss}e 1, D-47057 Duisburg, Germany}

\author[0000-0003-1665-5709]{Joel Green}
\affiliation{Space Telescope Science Institute, 3700 San Martin Drive, Baltimore, MD 21218, US}

\author[0000-0001-9443-0463]{Pamela Klaassen}
\affiliation{United Kingdom Astronomy Technology Centre, Edinburgh, GB}

\author[0000-0001-9030-1832]{Wafa Zakri}
\affiliation{Physical Sciences Department, Jazan University, Jazan, Saudi Arabia}

\author[0000-0002-2667-1676]{Nolan Habel}
\affiliation{Jet Propulsion Laboratory, California Institute of Technology, 4800 Oak Grove Drive, Pasadena, CA 91109, USA}

\author[0000-0001-7826-7934]{Nashanty Brunken}
\affiliation{Leiden Observatory, Universiteit Leiden, Leiden, Zuid-Holland, NL}

\author[0000-0002-5943-1222]{James Muzerolle}
\affiliation{Space Telescope Science Institute, 3700 San Martin Drive, Baltimore, MD 21218, US}

\author[0000-0002-7433-1035]{Katerina Slavicinska}
\affiliation{Leiden Observatory, Universiteit Leiden, Leiden, Zuid-Holland, NL}

\author[0000-0003-2300-8200]{Amelia M.\ Stutz}
\affiliation{Departamento de Astronom\'{i}a, Universidad de Concepci\'{o}n,Casilla 160-C, Concepci\'{o}n, Chile}

\author[0000-0002-9470-2358]{Lukasz Tychoniec}
\affiliation{Leiden Observatory, Universiteit Leiden, Leiden, Zuid-Holland, NL}

\author[0000-0002-0826-9261]{Scott Wolk}
\affiliation{Center for Astrophysics Harvard \& Smithsonian, Cambridge, MA, US}

\author[0000-0001-6144-4113]{Will R. M. Rocha}
\affiliation{Leiden Observatory, Universiteit Leiden, Leiden, Zuid-Holland, NL}

\author[0000-0002-3747-2496]{William J. Fischer}
\affiliation{Space Telescope Science Institute, 3700 San Martin Drive, Baltimore, MD 21218, US}

\begin{abstract}

Understanding the earliest stage of star and planet formation requires detailed observations to address the connection and interplay between the accretion, outflow and disk evolution.  We present results from the observations of the low luminosity ($L_\mathrm{bol}\sim~0.2~L_\odot$) and mass (M$_*\sim$\,0.15~M$_\odot$)  Class~0 protostar IRAS 16253$-$2429, conducted as part of the \textit{eDisk} ALMA large program and the JWST cycle-1 GO program \textit{IPA}. Observations reveal a wide hourglass-shaped continuum cavity traced in scattered light (at $\leq$~5~$\mu$m), with a brighter, extended northern side. We detect 15 pure rotational H$_2$ transitions (E$_\mathrm{up}$:~1015–21411~K), revealing a wide-angle molecular outflow.  {The outflow width (as traced in H$_2$~0-0~S(11)) at the protostellar location measures $\leq$35 au, slightly larger than the dust and Keplerian disk diameters ($\sim$30 au) but wider than the 20–23~au jet width in [Fe II].} {The opening angle narrows from 40–35\arcdeg{} for the low-J H$_2$ lines (up to S(5)) and the cold gas component (ALMA $^{12}$CO) to $\sim$28–19\arcdeg{} for the high-J H$_2$ lines (S(7)–S(11)).} Position-velocity diagrams of H$_2$ reveal higher velocities for higher E$_{up}$, ranging from ~12.5 km~s$^{-1}$ for H$_2$~0-0~S(1) and S(2)  to ~28.5 km~s$^{-1}$ for H$_2$~0-0~S(5)~and~S(7)  with respect to the mean flow velocity.  The nested excitation and velocity structure of the collimated jet and  wide angle wind suggest a magnetohydrodynamic wind as a likely launching mechanism, similar to the  findings in other protostars and Class II sources.  The lower velocity mm CO may be gas from the infalling envelope accelerated outwards by the wide angle wind along the cavity walls. 
\end{abstract}

\section{Introduction} \label{sec:intro}

The process of star formation is governed by the complex interplay between infall, accretion, and feedback mechanisms, which determine the final mass of the star and the initial conditions for planet formation \citep{2014prpl.conf..195D, 2014ApJ...790..128F,SF,2022MNRAS.515.4929G,2023ASPC..534..355F,2023ApJ...947...25H,2024ARA&A..62..203T,2024ARA&A..62...63H}. During the primary accretion phase (which includes most Class 0 and many Class I objects), protostars amass the bulk of their material due to high accretion rates \citep[e.g.,][]{2016ApJ...828...52W,2016ApJ...831...69M,2017ApJ...840...69F,2023ApJ...944...49F,2023arXiv230812689N}. It is also during the primary accretion phase that the protostellar disk is formed, and the initial conditions of planet formation are set \citep{2024ARA&A..62..203T}. 

Protostars, regardless of their mass (spanning three orders of magnitude), share a common structural blueprint \citep{2016A&ARv..24....6B,2024ARA&A..62..203T}, featuring a central protostar surrounded by a dense infalling envelope of gas and dust, and a circumstellar disk.  Collimated jets and wide-angle winds are also launched from the disk that carves cavities into the envelope, launching outflows of entrained gas and reducing the mass available to the nascent protostar  \citep{2020A&ARv..28....1L,2021NewAR..9301615R,2021ApJ...911..153H,2023ApJ...947...25H,2023ApJ...944..230S,2024MNRAS.533.3828D}. At the same time these jets and winds remove angular momentum from the disk, thereby driving accretion onto the central protostar \citep[e.g.,][]{2007prpl.conf..261S, 2016ARA&A..54..135H, 2019FrASS...6...54P, 2021NewAR..9301615R,2022MNRAS.512.2290T}.

\begin{figure*}
\centering
\includegraphics[width=1\linewidth]{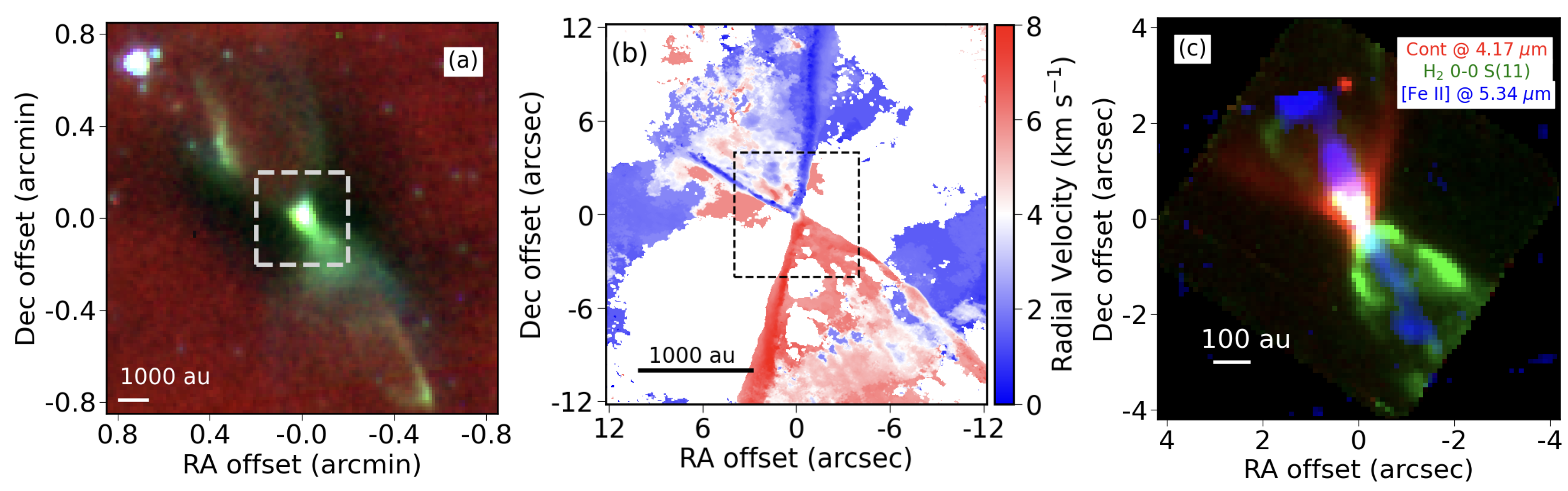}
\caption{(a) The Spitzer IRAC~three color (IRAC~3.6, 4.5 and 8 $\mu$m  are red, green, and blue, respectively) image of the IRAS 16253$-$2429 field. The field of view is $\sim$ 0.8\arcmin $\times$ 0.8\arcmin. The dashed gray square outlines the $\sim$ 24\arcsec\ $\times$ 24\arcsec{} field corresponding to the ALMA FOV in panel b. (b) Moment 1 (velocity) maps in the $^{12}$CO $J=2-1$ (robust =2) line from the ALMA eDisks observations. The dashed back square in the center outlines the $\sim$ 8\arcsec $\times$ 8\arcsec{} field centered on the JWST observations.  (c) A zoomed-in view of the NIRSpec and MIRI channel-1 Short (A) field with the red representing continuum emission at 4.17~$\mu$m, the green showing H$_2$ 0-0 S(11) at 4.18~$\mu$m and the blue depicting [Fe~II] at 5.34~$\mu$m. A scale bar is given on the bottom left corner for each of the images. }
\label{Figure1}
\end{figure*}

{Among the various proposed mechanisms for launching outflows, disk winds and X-winds are considered to be the leading contenders. Disk wind models have a characteristic nested velocity structure, where more collimated flows exhibit higher velocities \citep[e.g.,][]{1983ApJ...274..677P,1986ApJ...301..571P,1994ApJ...429..781S,1999A&A...343L..61C, 2007prpl.conf..261S, 2022A&A...668A..78D, 2023ASPC..534..567P,2023A&A...669A..81O,2025arXiv251221454F}. JWST observations of protostars and Class II disks also appear to find such nested outflows in atomic and H$_2$ emission from YSOs  \citep[e.g.,][Narang et al., in prep, Tyagi et al., in prep]{2024arXiv241018033P, 2024arXiv240916061C, 2024arXiv240319400D,2024A&A...687A..36T}; but also see \cite{2025arXiv250200394N}.   In contrast, X-winds have a similar velocity for the collimated and wide (opening) angle component, although recent models of X-winds generate a slower, wide angle component of shock compressed wind \citep{2020ApJ...905..116S,2023ApJ...945L...1S}. Recently \citet{2025arXiv251100515N} have demonstrated that photoevaporative wind models can also reproduce several key properties of the observed H$_2$ winds, such as their nested morphology and overall flux levels; however, these models generally underpredict the high-$J$ emission associated with more spatially extended H$_2$ structures.} 

A distinctive feature of disk winds is that they originate across a broad range of disk radii, from the inner disk regions around $\sim 0.1$ au to several 10s of au \citep{1983ApJ...274..677P,1986ApJ...301..571P,1993ApJ...410..218W,2000prpl.conf..759K,2007prpl.conf..277P,2019ApJ...874...90W,2023ASPC..534..567P,2023A&A...669A..81O}, with velocities ranging from a few km s$^{-1}$ to several tens of km s$^{-1}$. In contrast, X-winds are launched when stellar magnetic field lines connect with the disk at the co-rotation radius (around $\sim 0.1$ au), generating fast ($\sim 100$ km s$^{-1}$) MHD winds that spread widely \citep[e.g.,][]{1994ApJ...429..781S,2007prpl.conf..261S,2023ASPC..534..567P}. Another class of models to explain the outflows from protostars is the jet-driven bow-shock \citep[e.g.,][]{1993ApJ...414..230M,1993A&A...278..267R,2001ApJ...557..443O,2022A&A...664A.118R}. In this scenario, the jet propagates into the surrounding cloud and forms bow
shocks that accelerate and push the ambient gas, producing outflow shells surrounding the jet. Models invoking a pulsed jet create internal working surfaces in the jets that may entrain and accelerate gas inside the cavity, which may appear similar to a disk wind \citep{2022A&A...664A.118R,2023A&A...672A.116R}.

In this study, we examine the outflow  of IRAS 16253-2429, a Class 0 protostar in the nearby Ophiuchus molecular cloud \citep{2004A&A...426..171K,2006A&A...447..609S}.  The protostar has a bolometric luminosity of only $\sim$0.2~$L_\odot$ and a bolometric temperature of 42~K  (eHOPS database\footnote{https://irsa.ipac.caltech.edu/data/Herschel/eHOPS/overview.html}; \citealt{2023ApJS..266...32P}; also see \citealt{2023ApJ...951....8O}). At a distance of only $\sim$140~pc (adopted from  \citealt{2020A&A...633A..51Z}), IRAS~16253$-$2429 is the lowest luminosity source in the IPA sample \citep{2024ApJ...966...41F} and provides an excellent opportunity to study the jet, outflow, and cavity morphology for a class of (very) low-mass protostars in great spatial detail ($\sim$ 28 au) \citep{2023arXiv231014061N,2024ApJ...966...41F}. Such low luminosity protostars may represent  progenitors of very low-mass stars or substellar objects (close to the hydrogen-burning limit) stars very early in the accretion history, and/or low-mass protostars  undergoing a quiescent phase \citep{2004ApJS..154..396Y, 2005ApJ...633L.129B, 2006ApJ...649L..37B, 2006ApJ...640..391H, 2006ApJ...651..945D, 2007prpl.conf...17D, 2008ApJS..179..249D}. 

IRAS 16253-2429 has been extensively studied in the (sub)mm wavelength regime \citep[][]{2006A&A...447..609S,2011ApJ...740...45T, 2016ApJ...826...68H,2017ApJ...834..178Y,2019ApJ...871..100H,2023ApJ...954..101A,2025arXiv250919158S,2025arXiv251221454F}. The single dish observations from IRAM-30 telescope and APEX telescope \citep{2016ApJ...826...68H} have shown the CO outflow to be much larger than even the Spitzer IRAC emission spanning up to $\sim$3\arcmin~ ($\sim$0.12~pc) on the southern side. 

Recently, IRAS~16253$-$2429 was observed at high angular resolution ($\sim$0.\arcsec07 or 9.8 au for the 1.3~mm continuum) by the `Early Planet Formation in Embedded Disks eDisk' ALMA large program \citep{2023ApJ...951....8O,2023ApJ...954..101A}. These observations revealed a central protostar with a compact disk and a bipolar outflow detected in $^{12}$CO~$J=2-1$ (Figure \ref{Figure1}b). The blue-shifted emission is in the north-east lobe while the red-shifted emission is in the south-west lobe. 

The disk is resolved in the dust continuum at 1.3 mm. Similar to \cite{2024arXiv240509063Y} we define the dust disk radius as the 2$\sigma$ width of the major axis of the deconvolved 2D Gaussian function fitted to the 1.3 mm continuum image. The $2\sigma$ width of the major axis from fitting the 1.3 mm ALMA observations is $107\pm 2$~mas. At the distance of IRAS~16253$-$2429, this corresponds to a dust disk radius (2$\sigma$) of  $\sim$15~au. The disk mass, assuming an optically thin 1.3 mm dust continuum with a gas-to-dust mass ratio of 100, a disk temperature ranging between 20 and 27~K 
and different dust opacities, falls within the range of 1.4 to 2.1$\times 10^{-3}~M_\odot$ \citep{2023ApJ...954..101A}. The inclination angle of 64\arcdeg, was estimated based on disk fitting \citep{2023ApJ...954..101A}, which is consistent with the inclination angle between 60\arcdeg{}-65\arcdeg{} derived based on the modeling CO outflow by \cite{2017ApJ...834..178Y}.

\cite{2023ApJ...954..101A} also detected a Keplerian gas disk in $^{12}$CO with a radius of 16$\pm$3~au, similar to the dust disk as measured by the 1.3 mm continuum. By fitting the Keplerian  {rotation curve}, \cite{2023ApJ...954..101A} estimated the dynamical mass of the protostar to be in the range of 0.12$-$0.17~$M_\odot$. This means that IRAS~16253$-$2429 is forming a low-mass star and not a brown dwarf, contrary to previous speculations \citep{2016ApJ...826...68H}.  


The Spitzer Infrared Array Camera (IRAC) images of the source reveal a bipolar hourglass structure in the Northeast to Southwest direction (Figure \ref{Figure1}a). This hourglass structure traces the outflow cavities, along with shock features from the flow.  \cite{2010ApJ...720...64B} carried out spectral scan mapping of the protostar using the Spitzer IRS,  detecting six pure rotational H$_2$ lines (0-0~S(2) to 0-0~S(7)). However no fine structure lines were detected from the protostar with Spitzer/IRS.

Recent observations of IRAS 16253-2429 using the JWST Near Infrared Spectrograph (NIRSpec) integral field unit (IFU) and the Mid-InfraRed Instrument (MIRI) Medium Resolution Spectrometer (MRS) \citep{2024ApJ...966...41F, 2023arXiv231014061N} found that the bipolar hourglass-like structure (due to which the protostar gets its moniker of "wasp-waist" nebula) extends within the inner 7\arcsec\ and is traced in both scattered light as well as molecular H$_2$ lines (Figure \ref{Figure1}c). IRAS~16253$-$2429 also drives a bipolar jet seen in [Fe~II], Br-$\alpha$, and [Ne~II] \citep{2024ApJ...966...41F,2023arXiv231014061N}. {The MIRI MRS spectra of IRAS 16253-2429 shows weak H$_2$O but strong OH, indicating that H$_2$O is being photodissociated by UV photons from accretion shocks. Using the production rate of OH we can thus derive an accretion rate.  The accretion rate derived from the OH lines is $\dot{M}_{\rm acc} = 3 \pm 2.2 \times \ 10^{-10}~ M_{\odot}~ \mathrm{yr}^{-1}$ \citep{2025arXiv251215999W}}. The mass loss rate (based on the atomic jet) is $\leq 1.1 \times 10^{-10}~ M_{\odot}~\mathrm{yr}^{-1}$ \citep{2023arXiv231014061N}. Given the low mass accretion/ejection rates, the protostar appears to be going through a quiescent phase where the luminosity is dominated by the intrinsic luminosity of the central protostar.  Yet it is  driving a highly collimated atomic jet. 

\begin{table*}[]
\centering
\caption{JWST and ALMA Observations}
\begin{tabular}{|c|c|c|c|c|}
\hline
\multicolumn{5}{|c|}{}\\
\multicolumn{5}{|c|}{JWST}\\
\multicolumn{5}{|c|}{}\\\hline
Observation & RA$^*$ & Dec$^*$  & Start UT & End UT \\\hline
NIRSpec IRAS 16253-2429 & 16h28m21.62s & -24d36m24.16s & Jul 22, 2022 15:09:15 & Jul 22, 2022 17:59:43 \\
MIRI IRAS 16253-2429 & 16h28m21.62s & -24d36m24.16s  & Jul 23, 2022 04:04:41 & Jul 23, 2022 09:58:57 \\
MIRI background & 16h28m27.76s & -24d37m46.95s  & Jul 23, 2022 10:02:33 & Jul 23, 2022 11:35:19\\\hline
\multicolumn{5}{|c|}{}\\
\multicolumn{5}{|c|}{ALMA}\\
\multicolumn{5}{|c|}{}\\\hline
Projected baseline range & RA$^*$ & Dec$^*$ & \multicolumn{2}{c|}{Date} \\\hline
52--10540 m & 16h28m21.60s & -24d36m23.40s &\multicolumn{2}{c|}{2021 October 5, 26, 27, \& 28} \\
14--1290 m& 16h28m21.60s & -24d36m23.40s &\multicolumn{2}{c|}{2022 June 14 \& 15} \\\hline
\multicolumn{5}{c}{}\\
\end{tabular}
\label{log}
\tablecomments{{$^*$ The observations are centered at these coordinates.}}
\end{table*}

In this work, we present the results from joint JWST and ALMA observations of IRAS 16253-2429. The combination of ALMA and JWST is vital for understanding the morphology, kinematics, and excitation conditions in these low luminosity/mass stars, which could not have been done previously.  In section~2, we provide a brief overview of observations as well as the data reduction. Our results are reported in section~3, and discussed in detail in section~4. We summarize our findings in section~5.

\section{Observation and data reduction}
\subsection{JWST Observations}

IRAS~16253$-$2429 was observed as part of the JWST cycle-1 GO proposal Investigating Protostellar Accretion (IPA) (Program ID 1802, PI Tom Megeath; \citealt{2021jwst.prop.1802M}, \citealt{2023arXiv231014061N}; \citealt{2024ApJ...966...41F}, \citealt{2023arXiv231207807R}; \citealt{2024arXiv240407299N}; \citealt{2024arXiv240415399S}; \citealt{2024arXiv240204314B}; \citealt{2024arXiv241006697T}; \citealt{2026arXiv260109587F}). The main goal of the JWST observations was to perform spectral imaging by combining Integral Field Unit spectra of the Mid-InfraRed Instrument (MIRI) Medium Resolution Spectroscopy (MRS) \citep{2015PASP..127..584R,2015PASP..127..595W,2023PASP..135d8003W,2024arXiv240915435L} and the Near Infrared Spectrograph (NIRSpec) \citep{2022A&A...661A..80J,2022A&A...661A..82B} from 2.87 to 28~$\mu$m, covering the inner 5.\arcsec7~to 14.\arcsec6~region of the protostar at a resolution of $\sim$0.\arcsec2 ($\sim$28 au) to $\sim$ 1\arcsec{} ($\sim$140 au). The NIRSpec IFU and MIRI MRS observations were carried out nearly simultaneously. IRAS~16253$-$2429 was observed with a pointing center at RA = 16h28m21.62s  and Dec = -24d36m24.16s based on the 0.87~mm continuum peak from ALMA observations by \cite{2019ApJ...871..100H} (see Table \ref{log} for the observation log and \citealt{2023arXiv231014061N}).

For the NIRSpec IFU, we used the F290LP/G395M filter/grating combination and for MIRI we used all four channels in the MIRI Medium Resolution Spectroscopy (MRS) IFU mode. We mapped the protostar using a $2\times2$ mosaic with 10\% overlap and the 4-point dither mode.  A dedicated background with a similar observation setup (configurations and integration time) 118\arcsec~away from the target was also taken shortly 
 after the observation for MIRI MRS.

To reduce NIRSpec IFU data, we used the JWST pipeline version {1.9.5.dev7+gbf7d3c9b} \citep{bushouse_2023_7692609} and the JWST Calibration References Data System context version {jwst\_1069.pmap}. Details of the NIRSpec data reduction and the custom flagging routine are provided in \citep{2024ApJ...966...41F}. The MIRI MRS data reduction was carried out using the JWST pipeline version {1.11.3 \citep{bushouse_2023_8157276}} and the JWST Calibration References Data System context version {jwst\_1105.pmap} (see \citealt{2023arXiv231014061N} for further details). Upon examining the dedicated background IFU observation, we detected diffuse emission from the H$_2$ S(1) and H$_2$ S(2) molecular lines; therefore we also created data cubes without subtracting the dedicated background and used those for the analysis.

\begin{figure*}
\centering
\includegraphics[width=1\linewidth]{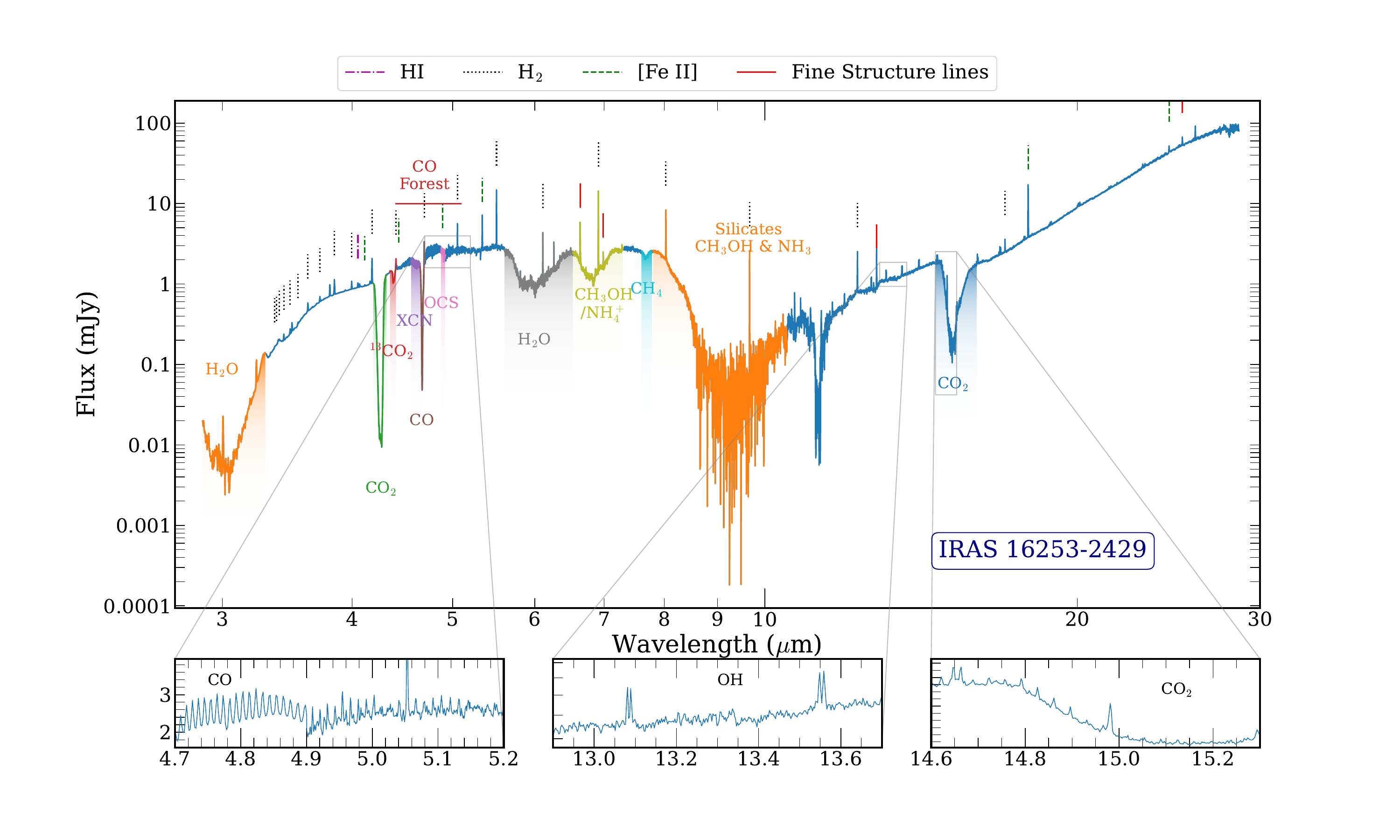}
\caption{The MIRI MRS and NIRSpec IFU complete spectrum from 2.87~$\mu$m to 28~$\mu$m extracted from the ALMA continuum position with an aperture  radius of 1\arcsec {centered at the 14~\micron{} protostellar position}. The offset plots show the zoomed-in view of some of the prominent molecular gas lines detected in IRAS 16253-2429. }
\label{Figure2}
\end{figure*}

We further refined the astrometry of our JWST observations, by aligning the MIRI parallel image (from the MIRI imager) to Gaia using Spitzer as an intermediary \citep[see][for details]{2024ApJ...966...41F}. This alignment between Gaia and the MIRI image carries an astrometric uncertainty of $\sigma$RA$_\mathrm{(Gaia-MIRI)}$=0.\arcsec05 and $\sigma$Dec$_\mathrm{(Gaia-MIRI)}$=~0.\arcsec04. We then correct the MIRI MRS observations using the MIRI parallel images. The NIRSpec IFU astrometry is then corrected using the overlapping spectral range between  MIRI MRS and NIRSpec IFU.

\begin{figure*}
\centering
\includegraphics[width=1.0\linewidth]{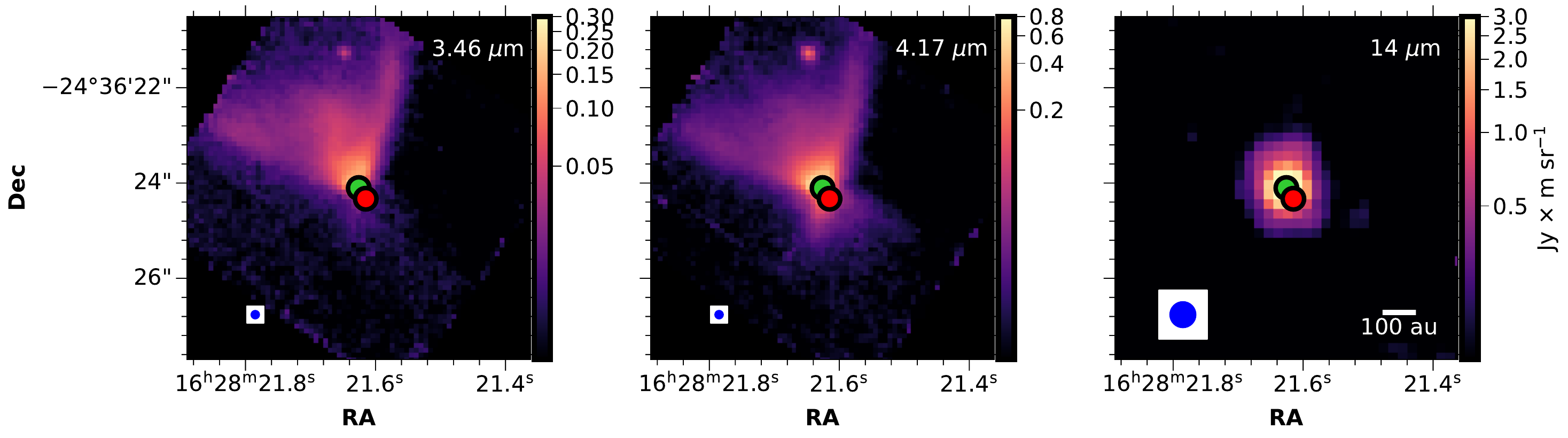}
\caption{The continuum morphology of IRAS~16253$-$2429 as a function of wavelength from NIRSpec IFU and MIRI MRS. All images are cropped to the same spatial scale.  The green marker is the MRS 14~\micron~ position, while the ALMA 1.3~mm position is shown as the red marker. A scale bar corresponding to 100~au is shown in the bottom right corner, while the beam size is shown in the bottom left corner as the blue circle.}
\label{Figure3}
\end{figure*}

\begin{figure*}
\centering
\includegraphics[width=1\linewidth]{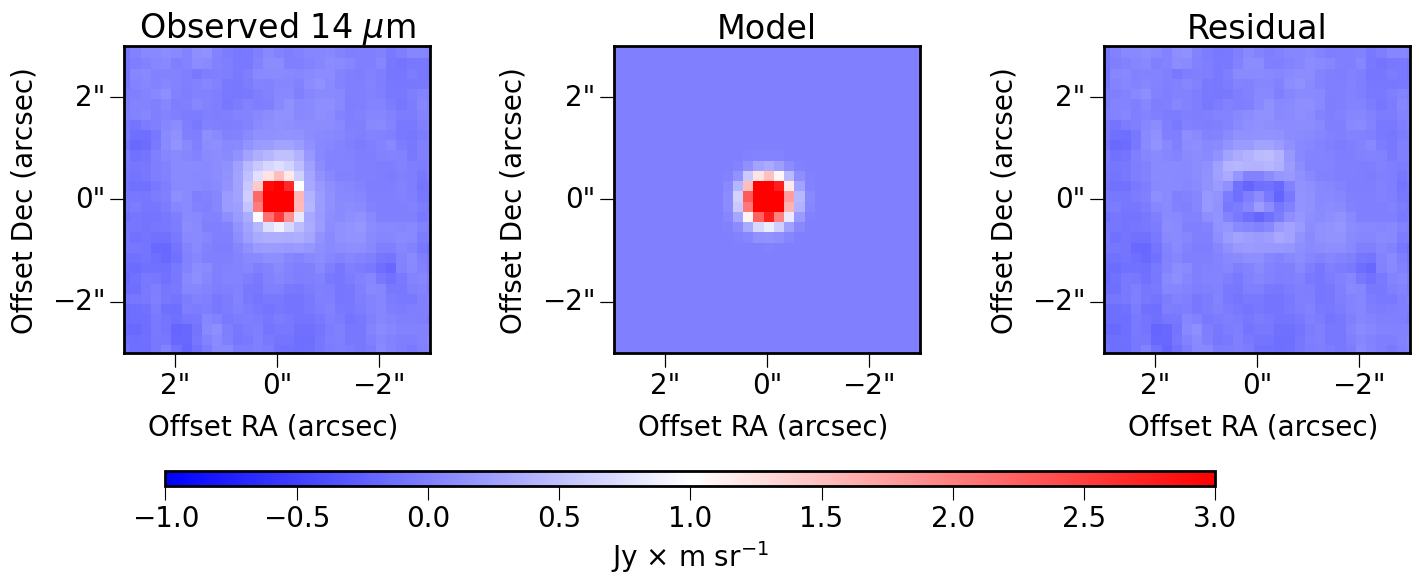}
\caption{(left){ The observed 14 \micron{} image, (center) the best-fit model and (right) the residual. All the three images are on the same spatial and intensity scale.} }
\label{2Dgauss}
\end{figure*}

\subsection{ALMA Observations}

IRAS~16253$-$2429 was observed with ALMA in Cycle~8 using the C-8 (with projected baseline lengths spanning from 52 to 10540~m) and C-5 (with projected baseline lengths ranging from 14 to 1290~m) antenna configurations as part of the ALMA large program 'Early Planet Formation in Embedded Disks' (eDisk) (2019.1.00261.L, 2019.A.00034.S) \citep{2023ApJ...951....8O}. The observations included CO ($J=2-1$) isotopologues , SO ($J_N = 6_5 - 5_4$), and other molecular lines, observed with Frequency Division Mode (FDM). Additionally, the 1.3~mm continuum emission was obtained from the line-free channels of the spectral windows, with a maximum window width of 1.875~GHz (see also \citealt{2023ApJ...951....8O} and \citealt{2023ApJ...954..101A}). {The continuum  image has a spatial resolution of 0.073\arcsec{} $\times$ 0.048\arcsec{} for a robust factor of 0.}  For the $^{12}$CO emission a velocity resolution of 0.32~km s$^{-1}$ at a angular resolution of 0.\arcsec17~$\times$ 0.\arcsec13 was achieved with a robust factor of 0.5.

\section{Results}

\subsection{Spectra towards the central source}

The near-simultaneous observations of IRAS~16253-2429 allowed for the combination of the MIRI MRS and NIRSpec IFU data to produce the complete spectrum of the target from 2.87~$\mu$m to 28~$\mu$m. We extracted the spectrum from the MIRI MRS 14~$\mu$m continuum position (see section 4.1) using an extraction aperture with a radius of 1\arcsec. In Figure \ref{Figure2}, we show the complete spectrum from IRAS~16253$-$2429. We detect several molecular H$_2$ lines, atomic and fine-structure lines, and the Brackett~$\alpha$ (Br~$\alpha$) HI line. While the H$_2$ lines trace the warm gas in the cavity from the protostar (Figure \ref{Figure1}c; also see \citealt{2024ApJ...966...41F}),  the fine structure lines of [Fe~II] and [Ne~II] along with the Br~$\alpha$ line trace a highly collimated, fast jet \citep[][also see Figure \ref{Figure1}c.]{2023arXiv231014061N}

In the 4-5~\micron~range of NIRSpec and MIRI channel-1 short, several CO fundamental rotation-vibration $v=1-0$ R and P branch emission lines with $J \leq$  40 are detected from the central region. These rotation-vibration CO lines may trace parts of the outflow from IRAS~16253$-$2429 
\citep{2023arXiv231207807R}.  We also detect CO$_2$ fundamental rotation-vibration lines in emission in the gas-phase (around 15~$\mu$m). On a careful investigation of the spectra from the protostellar position, we also detected pure rotational transitions of OH in emission with the ${}^2\Pi_{3/2}$ and ${}^2\Pi_{1/2}$ electronic configurations in the $\lambda~=~10-24~\micron$ range. We also detect very weak signatures of  H$_2$O emission in the $\lambda~=~5-7~\micron$ range leading  \citep{2025arXiv251215999W} to propose that much of the H$_2$O in the innermost region of IRAS~16253$-$2429 has been photodissociated into OH. In addition to gas lines, we detected various ice species, identified in Figure \ref{Figure2} using Table 1 in \citealt{2015ARA&A..53..541B}. Further details regarding the ice properties and modeling around IRAS~16253$-$2429 can be found in \cite{2024arXiv240204314B,2024arXiv240107901N}.

\begin{figure*}
\centering
\includegraphics[width=1\linewidth]{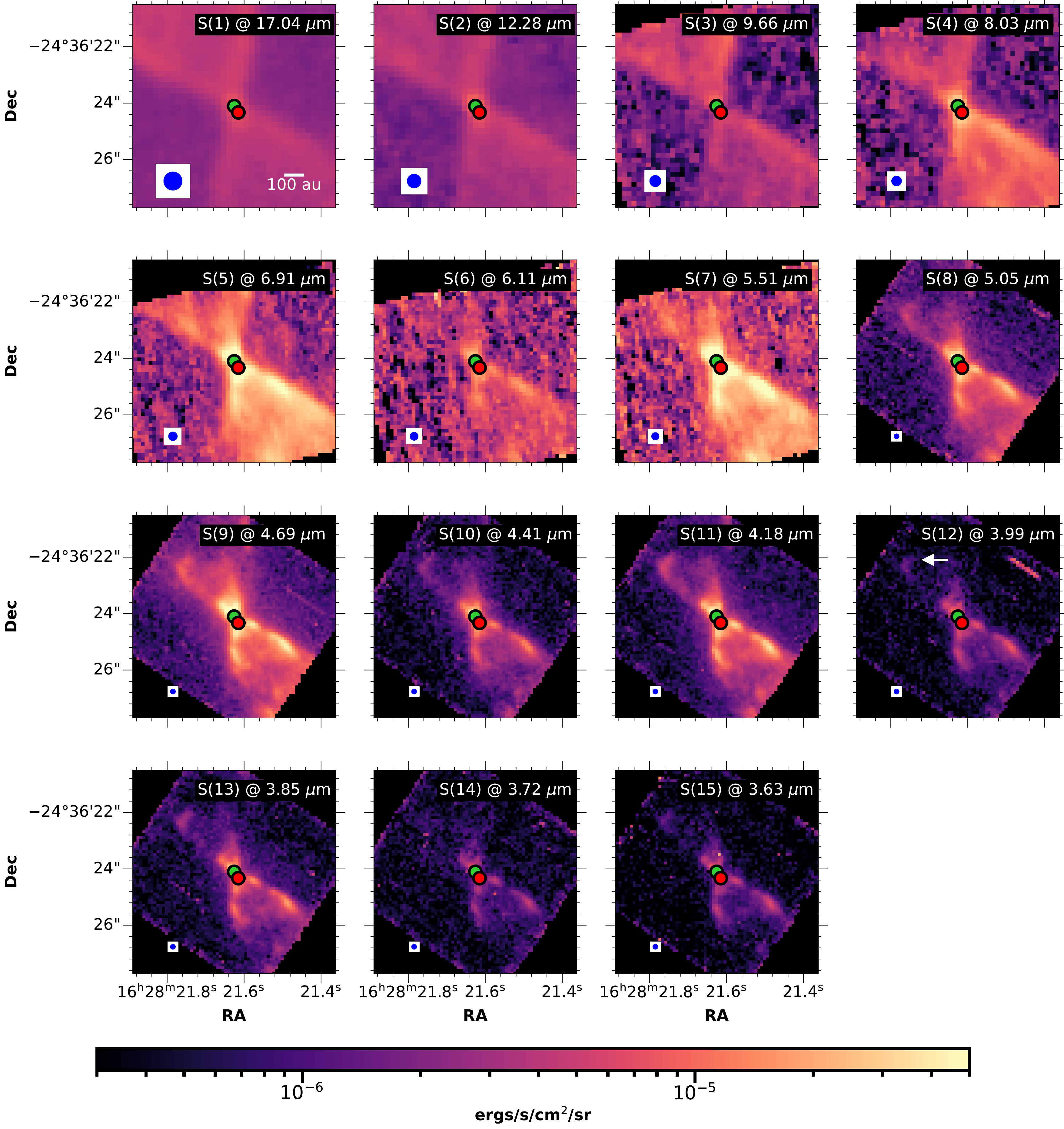}
\caption{The morphology of the H$_2$  0-0 S lines from S(1) to S(15). The  green  marker is the MRS 14~\micron~ position, while the ALMA 1.3~mm position is shown as the red marker. A scale bar corresponding to 100~au is shown in the bottom right corner (of the S(1) image), while the JWST PSF FWHM is shown in the bottom left corner as the blue circle. We also highlight the location of the spur in H$_2$ as a white arrow in the S(12) image. See Table \ref{TableH$_2$} for wavelength information for the H$_2$ transitions.}
\label{Figure4}
\end{figure*}

\begin{figure}
\centering
\includegraphics[width=1\linewidth]{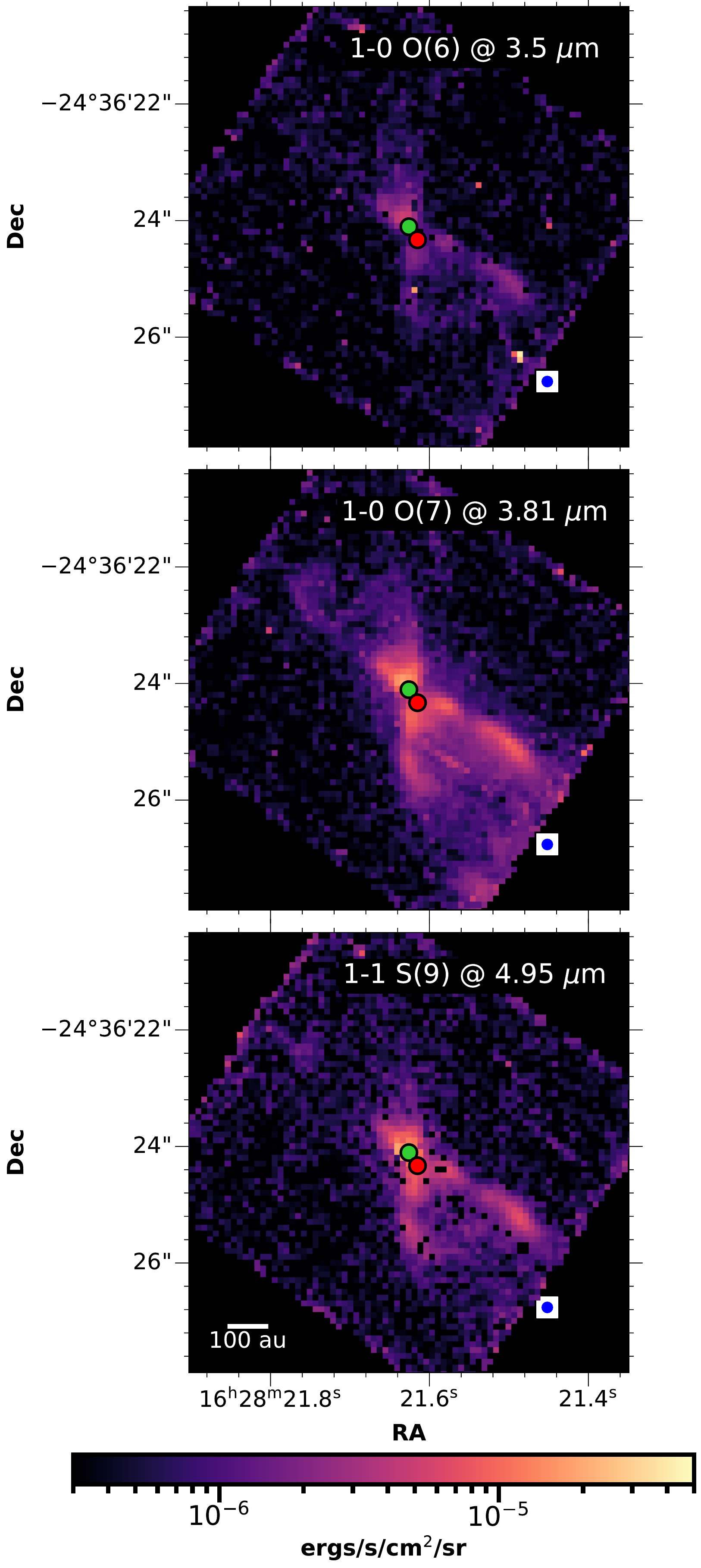}
\caption{The morphology of the high excitation H$_2$ lines detected towards IRAS~16253$-$2429. The green marker is the  MRS 14~\micron~ position, while the ALMA 1.3~mm position is shown as the red marker. A scale bar corresponding to 100~au is shown in the bottom left corner, while the JWST PSF FWHM is shown in the bottom right corner as the blue circle.}
\label{Figure5}
\end{figure}


\subsection{Scattered Light and Thermal Continuum}

JWST observations of IRAS~16253$-$2429 reveal ionic jets, molecular wide-angle outflows, and outflow cavity structures {down to a spatial resolution of 28~au} (with NIRSpec IFU). The collimated jet from the system has been extensively studied in \cite{2023arXiv231014061N} and \cite{2024ApJ...966...41F}; thus in this work we will focus on the cavity and molecular outflow morphology and their relationship to the disk and cavity morphology revealed by ALMA. 

In Figure \ref{Figure3}, we compare the continuum morphology of IRAS~16253$-$2429 across the NIRSpec IFU and MIRI MRS range {at 3.46 \micron{}, 4.17 \micron{} and 14 \micron{}. For the 14 \micron{} we have subtracted the median flux from the image to account for telescope background}. In the NIRSpec wavelength, the continuum traces the cavity in the scattered light and has a wide hourglass-like morphology (also see \citealt{2024ApJ...966...41F}). The continuum cavity is brighter on the northern side and extends further as compared to the southern side, consistent with the northern side (containing the blue-shifted outflow) being tilted towards us, this results in less extinction and more forward scattering \citep{2021ApJ...911..153H}. This is consistent with the higher extinction value found for the southern side of the protostellar cavity by \cite{2023arXiv231014061N}.  At longer wavelengths within the NIRSpec range, the southern cavity becomes increasingly visible due to reduced extinction.

At MIRI MRS wavelength $\gtrapprox$6~\micron{}, the continuum appears nearly as a point source. This can be attributed to the reduced scattering efficiency of dust grains in the mid-infrared \citep{2024RNAAS...8...68P}  in combination with the reduced sensitivity of the MRS observations compared to the NIRSpec IFU observations. {We fit the JWST 14 \micron{} continuum image using a 2D Gaussian to determine the protostellar position. The 2D Gaussian has an FWHM of 0.647\arcsec{}, which is slightly larger than the theoretical FWHM of $\sim$ 0.57\arcsec from \cite{2023AJ....166...45L}, suggesting that the source is marginally extended even at 14 \micron. The fit is shown in Figure \ref{2Dgauss}. The JWST 14 \micron{} continuum position is 16$^h$28$^m$21.626$^s$ -24$^d$36$^m$24.105$^s$, while the ALMA position from \cite{2023ApJ...954..101A} is 16$^h$28$^m$21.615$^s$ -24$^d$36$^m$24.33$^s$. }

\begin{figure*}
\centering
\includegraphics[width=0.4\linewidth]{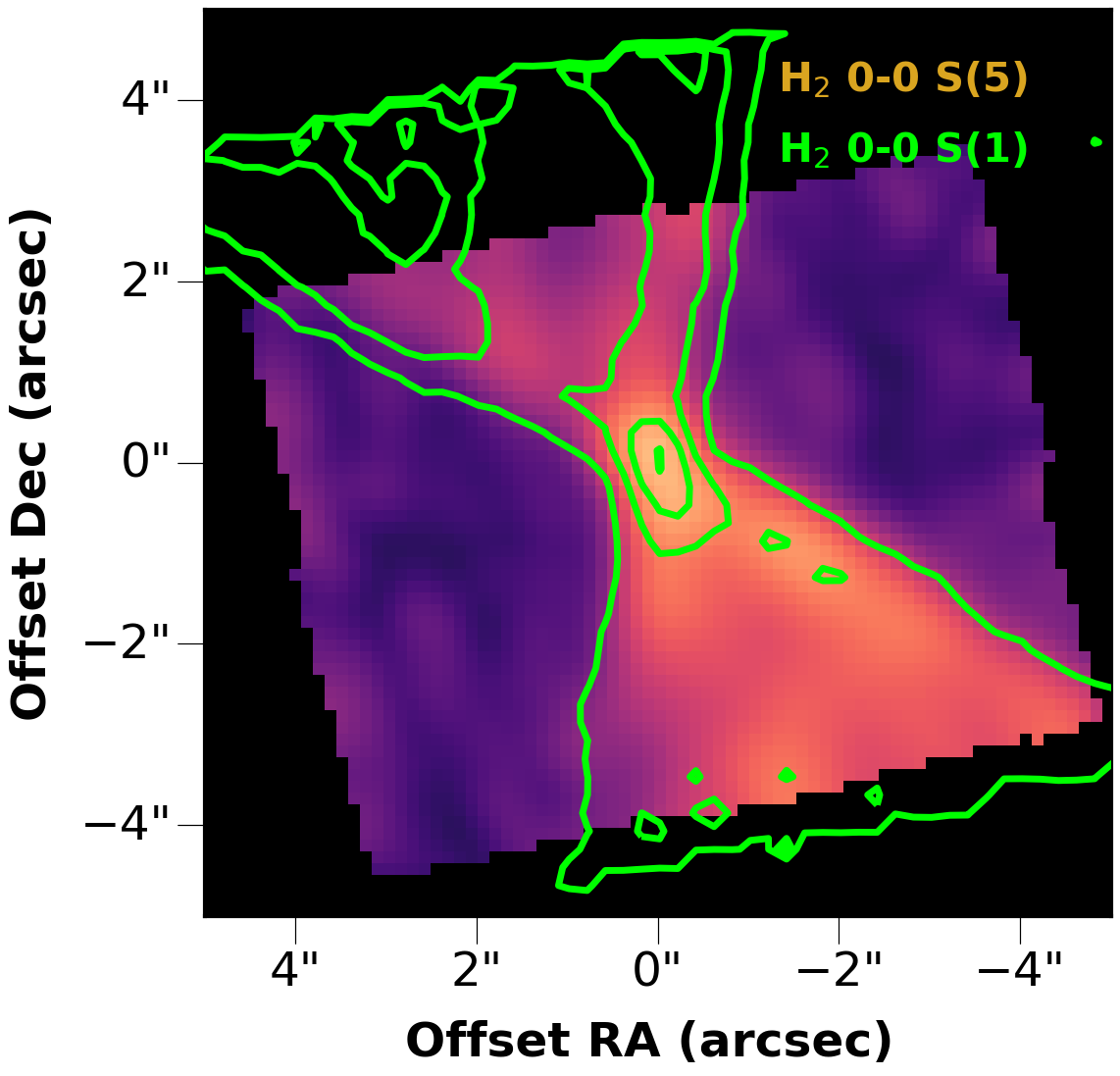}\includegraphics[width=0.4\linewidth]{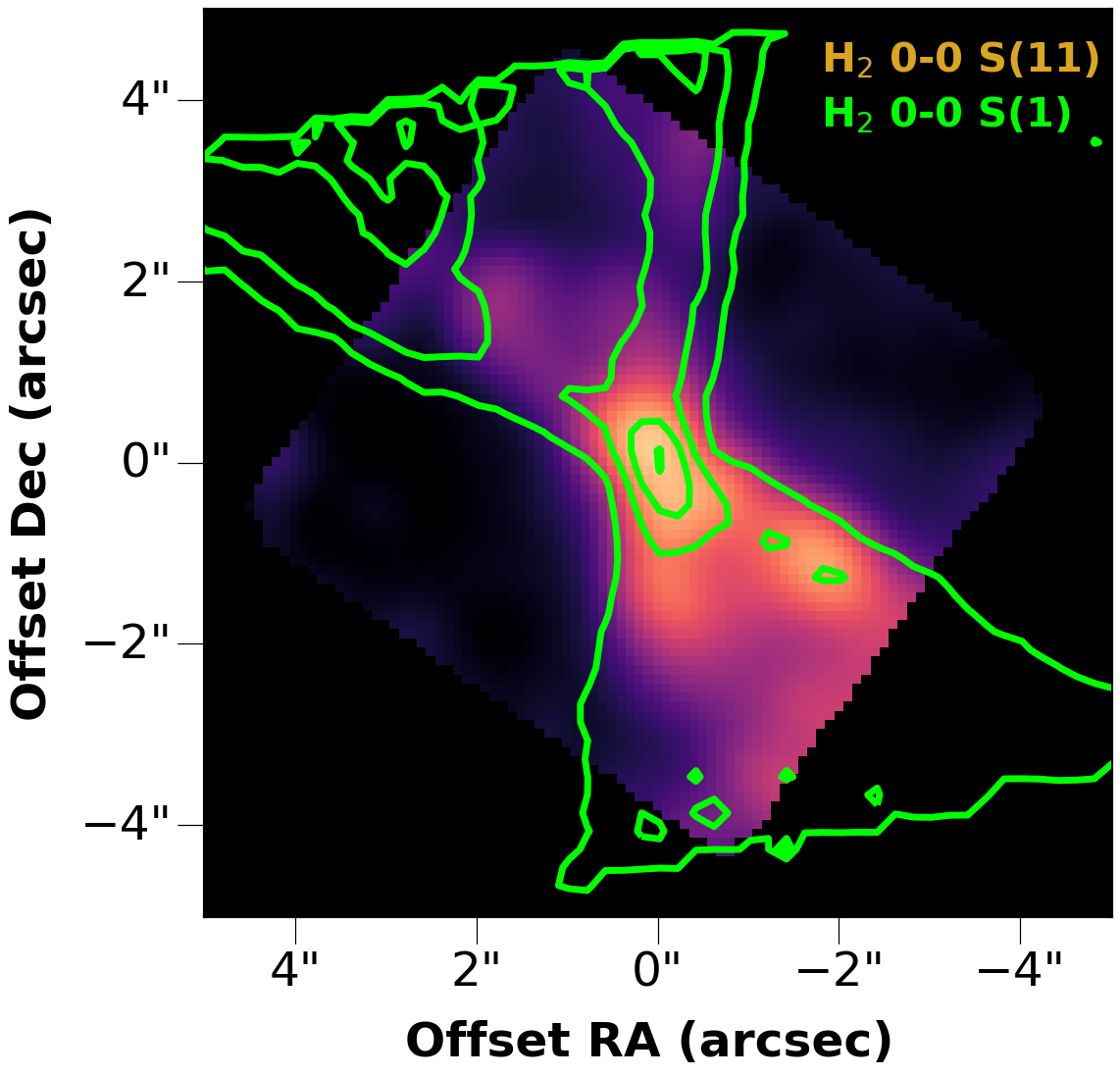}
\caption{A comparison of the width of the outflow, as traced in H$_2$ 0-0 S(1) (as lime contours) with the width of the outflow traced in  H$_2$ 0-0 S(5) and H$_2$ 0-0 S(11) (in color scale). The H$_2$ 0-0 S(5) and H$_2$ 0-0 S(11) {images have been convolved to the same resolution as S(1) and each panel is on a common pixel scale.} H$_2$ 0-0 S(1) contours are  40\%, 60\%, 80\% 99\% $\times$ peak line intensity of 7.8 $\times 10^{-6}$ erg s$^{-1}$ cm$^{2}$ sr$^{-1}$. }
\label{Figure6}
\end{figure*}

{We observe an offset of $\sim$0.\arcsec27 (38 au) ($\Delta$RA = RA$_\mathrm{JWST}$ - RA$_\mathrm{ALMA} \sim$ 0.\arcsec15 and $\Delta$Dec = Dec$_\mathrm{JWST}$ - Dec$_\mathrm{ALMA}\sim$ 0.\arcsec22) between the JWSR 14~\micron{} continuum position and the ALMA 1.3~mm continuum position} along the direction of the blueshifted jet; the offset is more than 3-4 times the estimated astrometric uncertainty of 0.064\arcsec \citep{2024ApJ...966...41F}. We further note that the ALMA continuum disk lies where the [Fe II] jet is narrowest (see Figure \ref{Figure8}). Even after correcting for the pointing offset for our IFU observations, this offset persists. 

Similar positional shifts ($\leq 1$\arcsec) between the positions of the outflow driving source at short wavelengths (optical/IR) and long wavelengths (cm/mm) have also been detected previously in other sources (e.g., HL Tau, HH 30, SVS13: \citealt{1995ApJ...452L.141W}; \citealt{1996ApJ...468L.103R}; \citealt{2022ApJ...930...91D}). 
\cite{2022ApJ...930...91D} argued that these  positional offsets could be due to the scattered light and thermal emission from illuminated, heated outflow cavity walls being detected while the central protostar is obscured. This effect tends to shift the short wavelength position in the direction of the blue-shifted lobe of the outflow (i.e., shifts the position toward the edge of the cavity closest to the observer, where the jet emerges outward from the cloud). This is similar to what we find in IRAS 16253-2429, as the IR position is displaced to the NE of the position of the embedded protostar (traced by the mm emission) along the blue-shifted jet/outflow. The positional offset, however,  does not have any major effects on the remainder of this study, but can be an important parameter for improving the radiative transfer model of the protostar.

\subsection{H$_2$ Emission in the Cavity}

In Figure \ref{Figure4}, we show the maps of the detected H$_2$ 0-0~S(J) lines towards IRAS~16253$-$2429 ranging from $J=$1 to $J=$15 (see Table \ref{TableH$_2$} for spectroscopic details). At every pixel, a local (line-free) continuum was subtracted, then a Gaussian fit to the  continuum subtracted line profile was made, and the integrated flux (from the area of the Gaussian) at each pixel was computed.    However, the signal-to-noise ratio (S/N) for H$_2$ 0-0~S(3) and H$_2$ 0-0~S(6) is notably lower. This is because the H$_2$ 0-0~S(3) line falls within the 10~$\mu$m silicate feature, while the H$_2$ 0-0~S(6) line is situated within the 6~$\mu$m H$_2$O ice feature. All the H$_2$ 0-0~S(J) lines trace the protostellar molecular outflow and exhibit an hourglass-like shape similar to the $^{12}$CO outflow detected at (sub)millimeter wavelengths \citep{2016ApJ...826...68H,2017ApJ...834..178Y,2019ApJ...871..100H,2023ApJ...954..101A}. Based on the velocities measured in Section~\ref{sec:velocity}, we interpret the H$_2$ as emission from a warm, molecular outflow. 
Furthermore, we observe weak extended emission for H$_2$ lines beyond the boundaries of the cavity, as delineated by the NIRSpec continuum. This may be ambient emission from the surrounding molecular cloud (also see Tyagi et al.,  in prep).

\begin{figure*}
\centering
\includegraphics[width=0.33\linewidth]{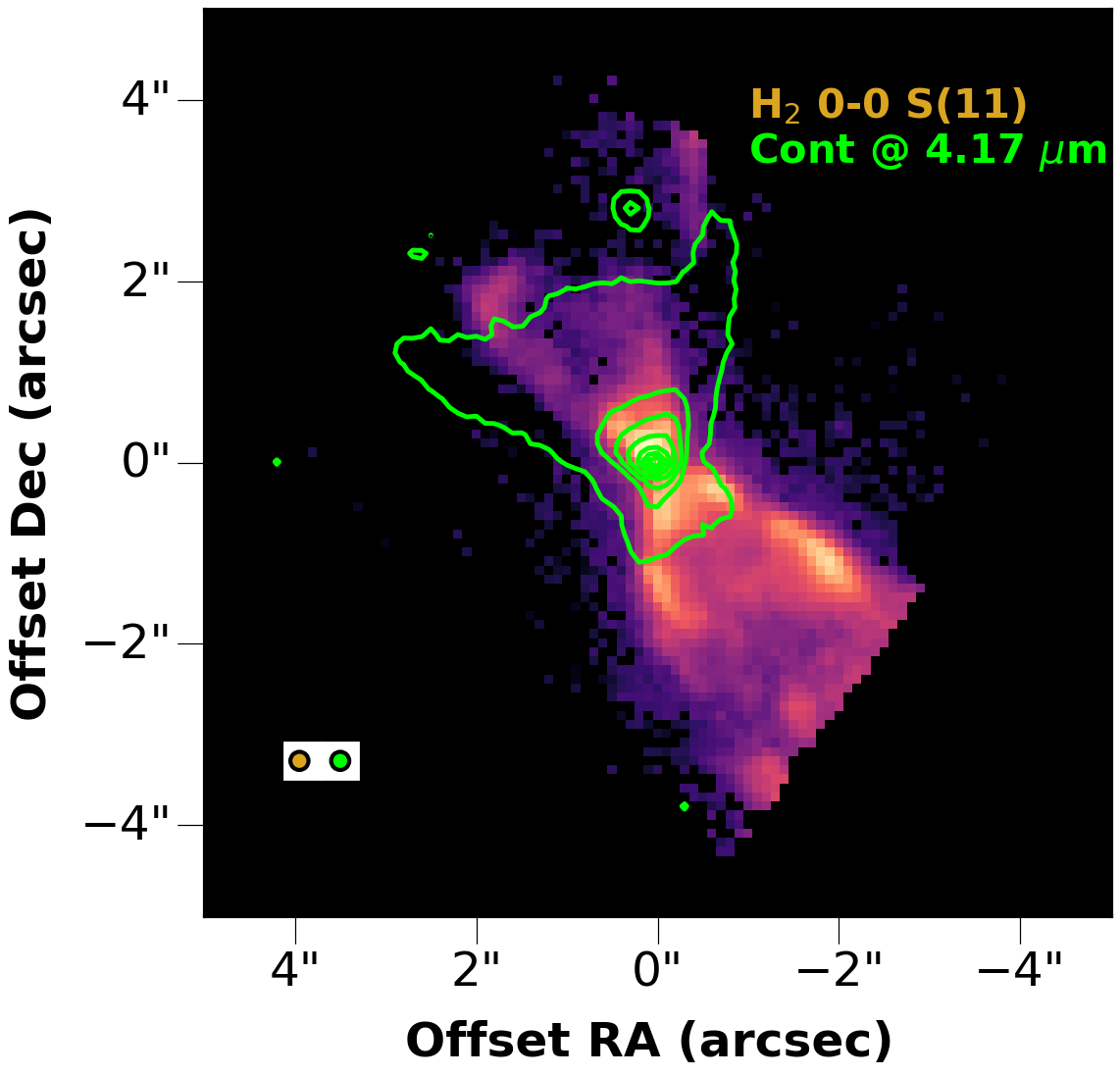}\includegraphics[width=0.33\linewidth]{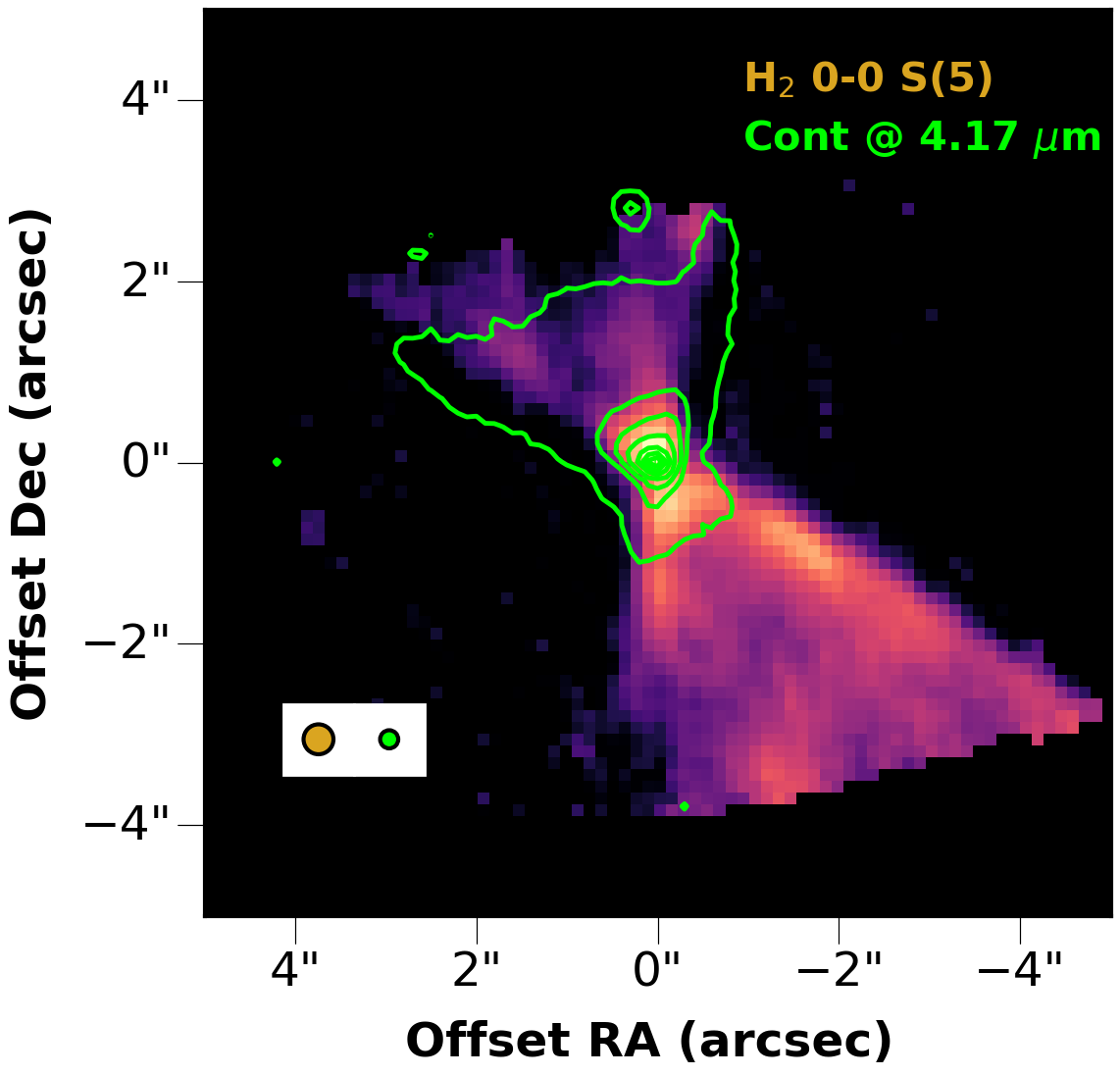}
\includegraphics[width=0.33\linewidth]{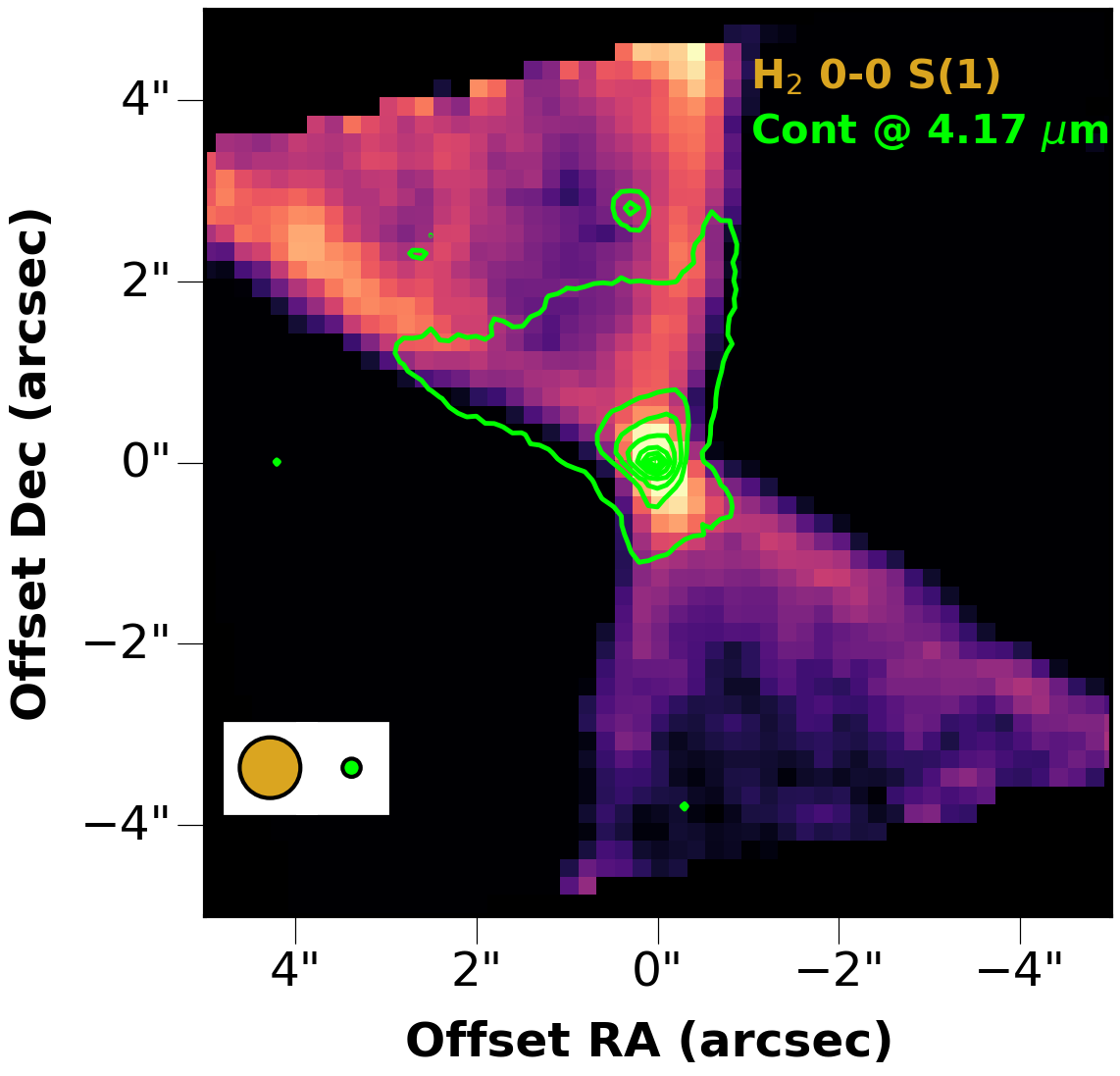}
\caption{A comparison of the cavity seen in scattered light in the continuum at 4.17~$\mu$m (lime contours) and the H$_2$  0-0~S lines from NIRSpec and MIRI wavelength range (color scale). The  NIRSpec continuum contours are  1\%, 5\%, 10\%, 20\%, 40\%, 60\%, 80\% 99\% $\times$ peak continuum intensity of 2.09 MJy $\mu$m sr$^{-1}$. {The JWST PSF is shown in the bottom left corner (lime for the NIRSpec continuum and khaki for the H$_2$ lines)} }
\label{Figure7}
\end{figure*}

In addition to the H$_2$ 0-0~S(J) lines, extended emission from transitions of H$_2$ 1-0~O(J) and H$_2$ 1-1~S(J) lines are also detected. These are rotational line in the upper $v$ =1 state, that can be used to compute the vibrational temperature. 
In Figure \ref{Figure5} we show the bright ro-vibration lines detected in NIRSpec IFU.  Despite the low S/N of these lines, their morphology is similar to that of the high-excitation (J $\geq$8) H$_2$ 0-0~S(J) lines. However, no H$_2$ 1-0~Q(J) transitions were detected from the protostar.

The H$_2$ outflow emission from IRAS~16253$-$2429 is not symmetric between the northern and southern cavities. We find that for higher-J transitions of H$_2$ the southern cavity (red-shifted) is much brighter than the northern side (opposite to what we found for the continuum emission). This suggests that the excitation of the H$_2$ is different between the two lobes; this will be explored in a subsequent paper.  The low transition lines fill the cavities with an enhancement towards the wall of the cavity indicating limb brightening due to shocks along the cavity wall.  For higher transitions, the flow is narrower in the northern cavity.  This cavity also shows a spur (Figure \ref{Figure4}) toward the jet \citep[also see][]{2024ApJ...966...41F}. In the southern cavity, bright emission is seen along the cavity edge. On the northern cavity, only a small length along the  NW wall shows emission in the higher transitions.  

\begin{deluxetable}{cllcc}
    \tablewidth{0pt}
    \tablecaption{The H$_2$  lines analyzed in this work. The wavelength, Einstein A coefficient $A_{ul}$, and upper state energy are from \cite{2022JQSRT.27707949G}. The symbol $^*$ indicates that the lines are not pure rotational transitions.  }
    \label{TableH$_2$}
    \tablehead{
    \colhead{Wavelength} &  \colhead{Name} &   \colhead{$A_{ul}$} & \colhead{$E_{up}$}  \\
    \colhead{$\micron$} & \colhead{} &  \colhead{s$^{-1}$}   &  \colhead{K} \\
    }
    \decimals
    \startdata
17.035	 & 	H$_2$ 0-0 S(1)	 		 & 	4.8 $\times 10^{-10}$ 	 & 		1015		 \\
12.279	 & 	H$_2$ 0-0 S(2)	 		 & 	  2.8 $\times  10^{-9}$	 & 		1682		 \\
9.665	 & 	H$_2$ 0-0 S(3)	   		 & 	 9.8 $\times 10^{-9}$ 	 & 		2504		 \\
8.025	 & 	H$_2$ 0-0 S(4)           & 	 2.6 $\times 10^{-8}$ 	 & 		3475		 \\ 
6.910	 & 	H$_2$ 0-0 S(5)	 	  		 & 	 5.9 $\times 10^{-8}$ 	 & 		4586		 \\
6.109	 & 	H$_2$ 0-0 S(6)	 	 		 & 	 1.1 $\times 10^{-7}$ 	 & 		5830	 \\ 
5.511	 & 	H$_2$ 0-0 S(7)	 	 		 & 	2.0 $\times 10^{-7}$ 	 & 		7197		 \\
5.053	 & 	H$_2$ 0-0 S(8)	 	 		 & 	  3.2 $\times  10^{-7}$	 & 		8677		 \\
4.954$^*$    & H$_2$ 1-1 S(9)                &  4.3 $\times 10^{-7}$ 	 & 	15722		 \\
4.695	 & 	H$_2$ 0-0 S(9)	 	 & 		 	 4.9 $\times 10^{-7}$ 	 & 		10261		 \\
4.41	 & 	H$_2$ 0-0 S(10)	 	  	 & 	 7.0 $\times 10^{-7}$ 	 & 		11940		 \\ 
4.181	 & 	H$_2$ 0-0 S(11)	 	  		 & 	 9.6 $\times 10^{-7}$ 	 & 		13703		 \\
3.996	 & 	H$_2$ 0-0 S(12)	 	 		 & 	 1.3 $\times 10^{-6}$ 	 & 		15540 \\
3.846	 & 	H$_2$ 0-0 S(13)	 	 		 & 	 1.6 $\times 10^{-6}$ 	 & 		17444 \\
3.807$^*$	 & 	H$_2$ 1-0 O(7)	 	 		 & 	 1.1 $\times 10^{-7}$ 	 & 		8365 \\
3.724	 & 	H$_2$ 0-0 S(14)	 	 		 & 	 2.0 $\times 10^{-6}$ 	 & 		19403 \\
3.626	 & 	H$_2$ 0-0 S(15)	 	 		 & 	 2.4 $\times 10^{-6}$ 	 & 		21411 \\
3.501$^*$	 & 	H$_2$ 1-0 O(6)	 	 		 & 	 1.5 $\times 10^{-7}$ 	 & 		7584 \\
\enddata
\end{deluxetable}

When comparing the morphology of various transitions of H$_2$ (Figure \ref{Figure4}), it is evident that the higher energy transitions of H$_2$ appear to be more collimated and delineate a narrower outflow compared to lower energy transitions. However, it is important to note that the various H$_2$ lines have different resolutions, ranging from 0.\arcsec2 to $\sim$0.\arcsec67{} \citep{2024ApJ...966...41F,2023AJ....166...45L}. Therefore, to ensure a fair comparison, we convolved the H$_2$ emission lines to the common spatial resolution of the 0-0 S(1) line. In Figure \ref{Figure6}, we superimposed the convolved maps of H$_2$ 0-0~S(5) and S(11) (chosen for their high S/N) on top of the  H$_2$ 0-0 S(1) line (at 17.035 \micron). The outflow traced in 0-0 S(1) appears to be broader than the outflow observed in S(5) and S(11), even after convolution (to a common spatial resolution);  this effect is more pronounced on the northern side. 

\begin{figure*}
\centering
\includegraphics[width=0.4\linewidth]{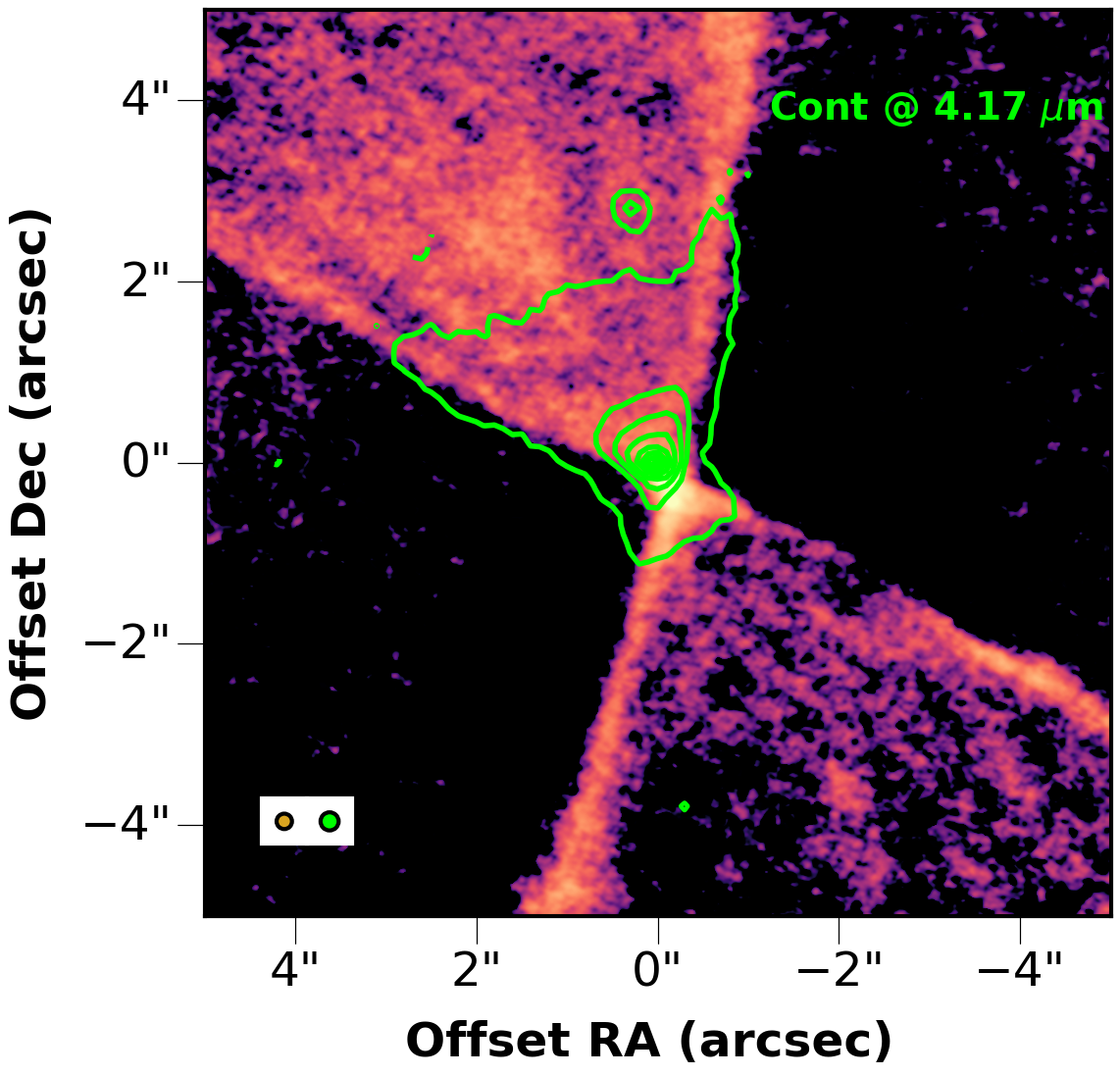}\includegraphics[width=0.4\linewidth]{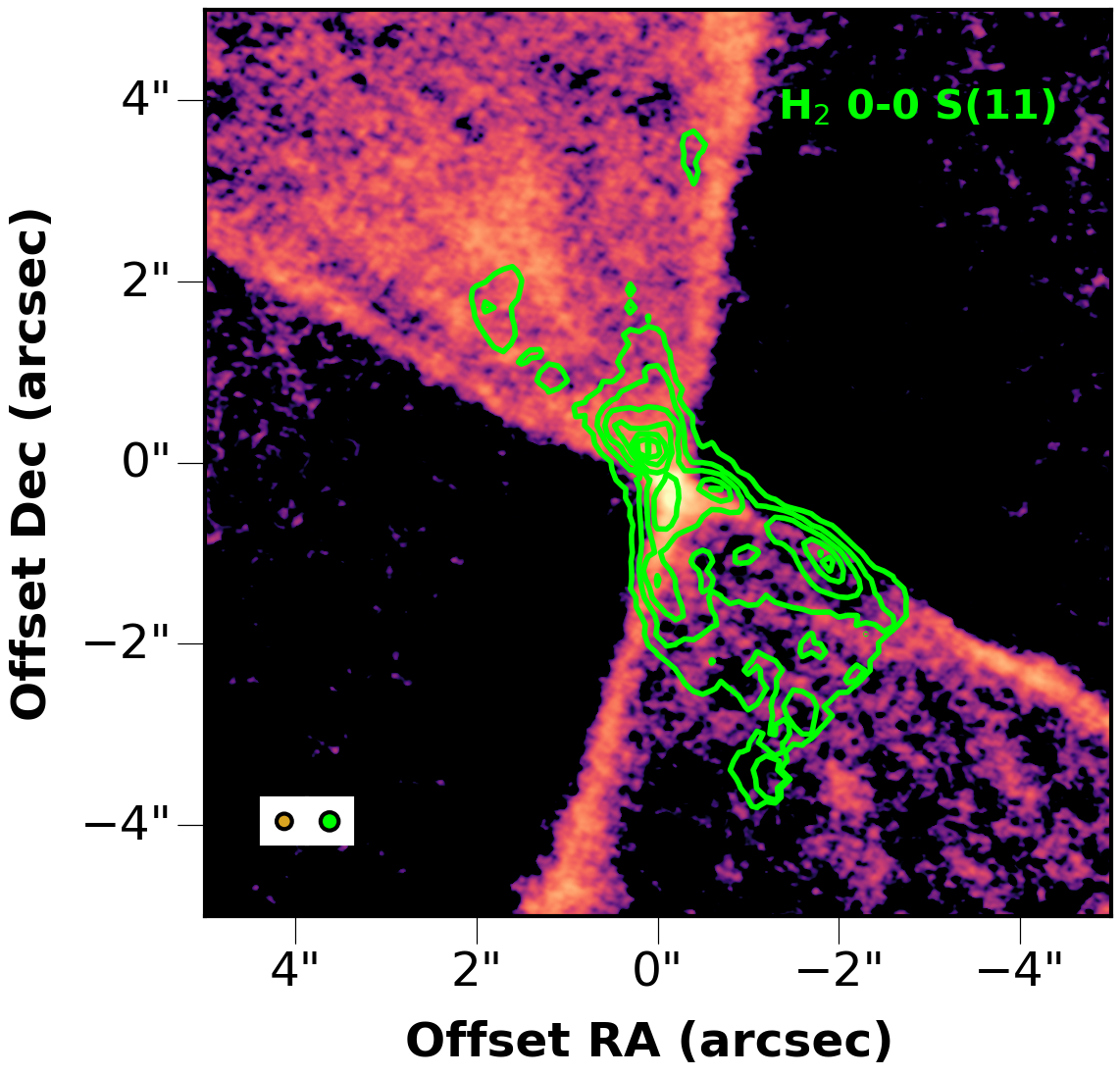}

\includegraphics[width=0.4\linewidth]{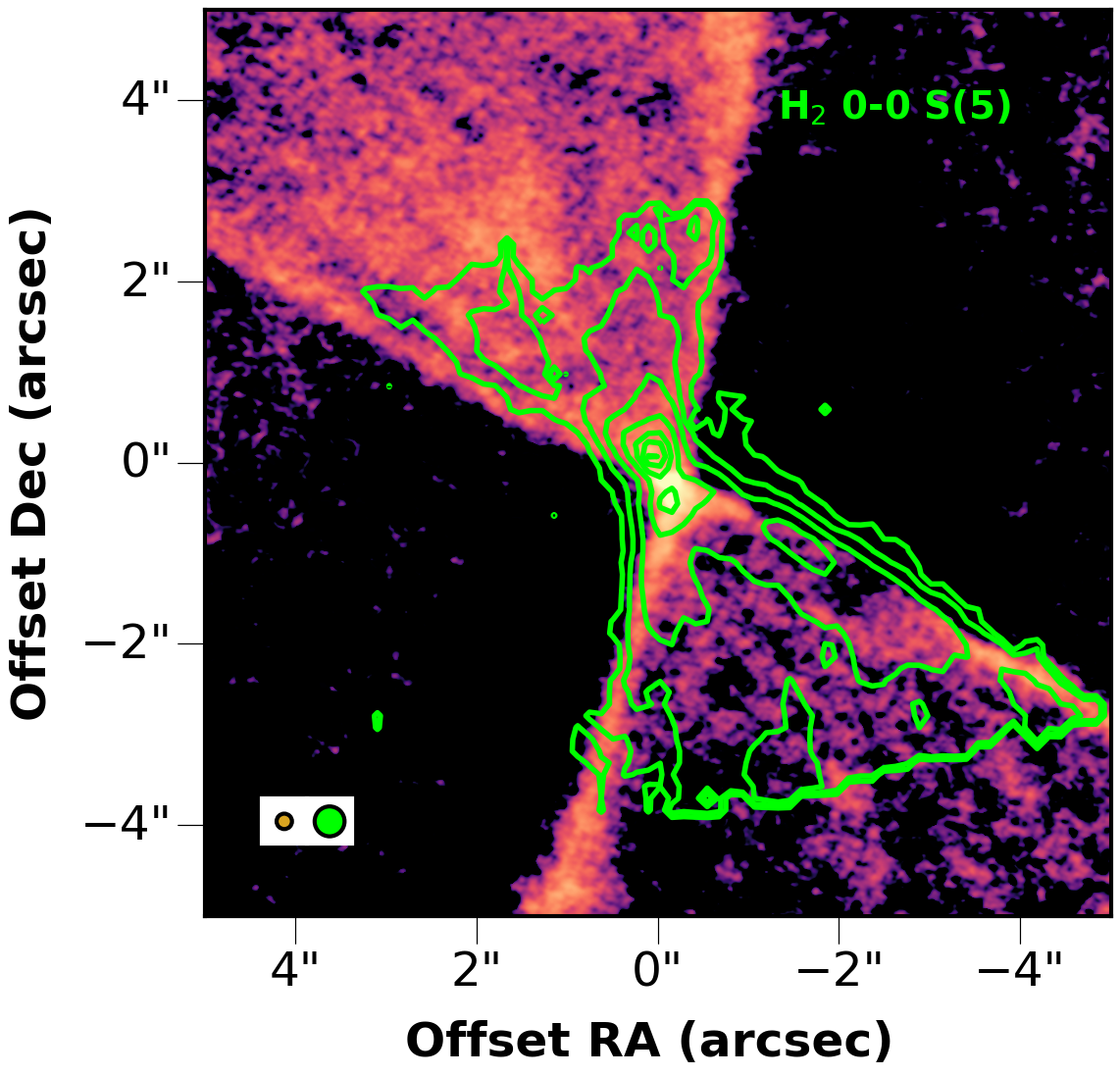}\includegraphics[width=0.4\linewidth]{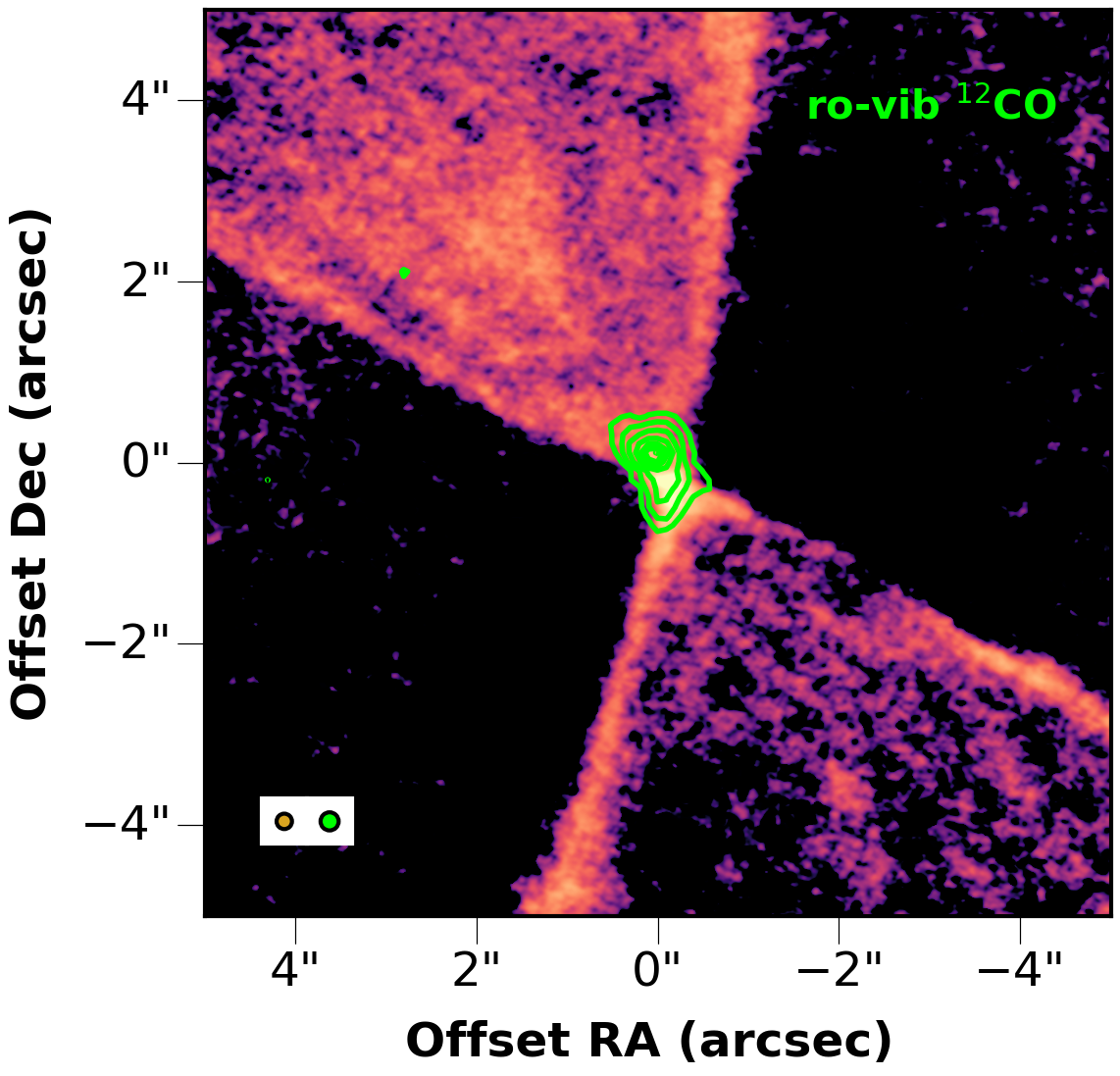}
\caption{The ALMA integrated intensity CO 2--1 (robust=0.5)  emission map (color scale) with the JWST continuum at 4.17~\micron{}, H$_2$ 0-0~S(11), H$_2$ 0-0~S(5)  and the ro-vibrational $^{12}$CO lines as lime contours. The contours are 5\%, 10\%, 20\%, 40\%, 60\%, 80\% 99\% $\times$ peak (line) intensity for the gas lines, while the outer-most contour for the continuum is at 1\% level. {The JWST PSF (lime circle) and the maximum ALMA beam (khaki circle) is shown in the bottom left corner.} }
\label{Figure8}
\end{figure*}

\subsection{H$_2$ Emission in context of Scattered Light Continuum and the ALMA $^{12}$CO Outflow}

We next analyzed the width of the H$_2$ emission and compared it with the width of the scattered light cavity observed using NIRSpec. In Figure \ref{Figure7}, we overlay the NIRSpec continuum on the H$_2$ emission lines. We examined how the morphology of the H$_2$ lines corresponds to the 4.17~$\mu$m continuum cavity, focusing on three high-S/N lines: 0-0~S(11), 0-0~S(5) and 0-0 S(1). The 0-0~S(11) line has a resolution of $\sim$0.\arcsec2 \citep{2024ApJ...966...41F}, comparable to the resolution of the continuum cavity, while 0-0~S(5) has a slightly lower resolution of $\sim$0.\arcsec33{} and 0-0 S(1) has a resolution of $\sim$0.\arcsec67{} \citep{2023AJ....166...45L}. The H$_2$ emission is narrower than the 1\% contour of the scattered light continuum tracing the cavity wall on the northern side, not only for the higher energy transitions but also for S(1). 

\cite{2023ApJ...954..101A} recently studied the molecular outflow from IRAS~16253$-$2429 at high resolution. These observations revealed a wide-angle bipolar outflow detected in $^{12}$CO $J=$ 2--1 (also see Figure \ref{Figure1}b), but no collimated molecular jet was detected. The bulk of the $^{12}$CO emission has a velocity within $\pm 4$km~s$^{-1}$ of the systemic V$_\mathrm{LSR}$ of the protostar of $\sim$ 4~km~s$^{-1}$. In Figure \ref{Figure8} we show the ALMA $^{12}$CO $J=$ 2--1 outflow (integrated from -1 to 10 km~s$^{-1}$, robust =0.5) superimposed on the NIRSpec continuum, H$_2$ lines, and the ro-vibrational $^{12}$CO lines detected in NIRSpec \citep{2023arXiv231207807R}. The resolution of the $^{12}$CO $J=$2--1 (0.\arcsec17 $\times$ 0.\arcsec13) is better than the NIRSpec resolution of $\sim$0.\arcsec2 \citep{2024ApJ...966...41F} and the MIRI resolution (0.\arcsec33)  at the wavelength of the H$_2$ S(5) line.

The scattered light cavity (traced by the NIRSpec continuum) exhibits a width slightly wider tham that of the ALMA $^{12}$CO $J=$ 2--1 outflow across the northern outflow. The 1\% contour of the continuum extends slightly beyond the CO emission, which could also be due to a combination of angular resolution and penetration of scatted light beyond the cavity wall. The low-$J$ H$_2$  (with E$_{up}\leq5000$ K) emission also has a similar width to the  ALMA $^{12}$CO $J=$ 2--1 outflow in the northern and southern cavities. The high-$J$ (with E$_{up}>5000$ K) H$_2$ emission is narrower in the northern part of the outflow but extends across the entire southern ALMA 12 CO (2-1) outflow, extending to the cavity walls. 
The ro-vibrational $^{12}$CO emission is concentrated at the central protostar and appears to trace the neck of the cold molecular $^{12}$CO $J=$2--1 from the outflow. 

\begin{figure*}
\centering
\includegraphics[width=0.5\linewidth]{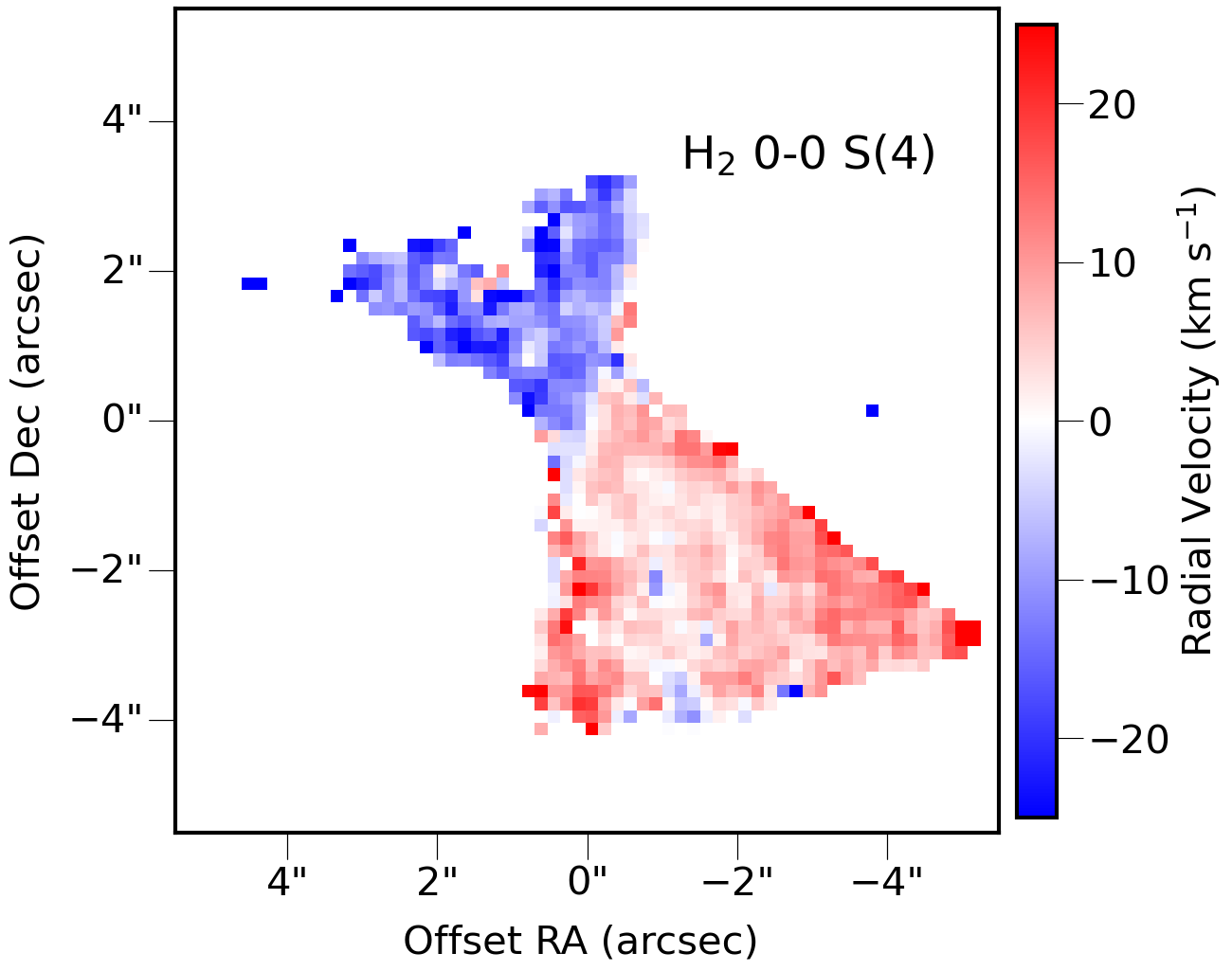}\includegraphics[width=0.5\linewidth]{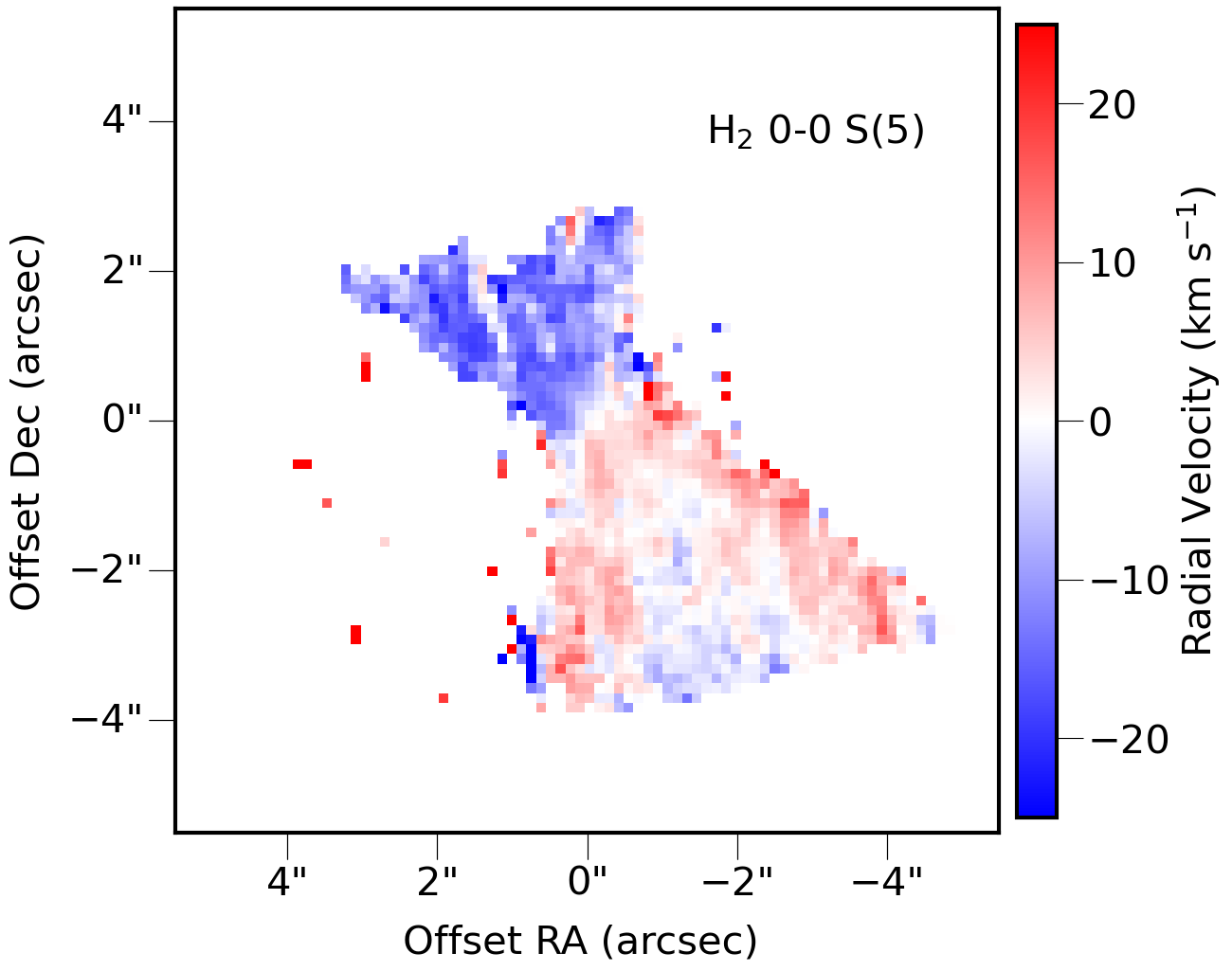}
\caption{Velocity maps of the H$_2$ 0-0~S(4) and H$_2$ 0-0~S(5) lines at 8.02 \micron{} and 6.91 \micron.}
\label{Figure12}
\end{figure*}

\subsection{Velocity Structure of the H{$_2$} flow}\label{sec:velocity}

To investigate the velocity structure within the H$_2$ outflow, we generated velocity maps of the H$_2$ 0-0~S(4) and H$_2$ 0-0~S(5) lines at 8.02~\micron{} and 6.91~\micron{}, respectively. These lines have high S/N as well as high spectral and spatial resolution (as well as being a close pair of ortho and para H$_2$ lines).  The velocity maps serve as a valuable diagnostic tool for discerning the kinematic structure within the jets and outflows. During the Gaussian fitting procedure used to generate line maps, one of the extracted parameters is the line center. By measuring shifts in the line center, we can derive velocity shifts, which in turn are used to construct velocity maps. To make the velocity maps we have masked pixels with emission $<$ 6~$\sigma$. We further correct for any systematic offset in the velocity by  computing the mean velocity of the velocity maps and subtract that out from each pixel.

In Figure \ref{Figure12}, we show the velocity maps of the H$_2$ 0-0~S(4) and H$_2$ 0-0~S(5) lines.  We find that the maximum (blue-shifted or red-shifted) velocity of the outflow (as observed) is $\sim$ 20~km~s$^{-1}$ (after correcting for systemic velocity) and if we correct for inclination (with inclination angle of 64\arcdeg) the maximum velocity is $\sim$ 45~km~s$^{-1}$.   We find that similar to the $^{12}$CO $J=$2--1 line seen in ALMA, the northern outflow is blue-shifted, and the southern outflow is red-shifted. This shows that the H$_2$ emission is tracing a bipolar outflow in the central cavity surrounding the central jet. The H$_2$ flow from the protostar is however much faster than the  $^{12}$CO $J=$2--1 (see Figure \ref{Figure1}b and \citealt{2023ApJ...954..101A}), which has a radial velocity of 1.5-2.5 km s$^{-1}$ and after correcting for inclination has a radial velocity of 3.4 to 5.7 km s$^{-1}$. The H$_2$ velocity in turn is much slower than the jet velocity of  169 $\pm$ 15~km~s$^{-1}$ \citep{2023arXiv231014061N}. We further note that at the protostellar location part of the blue (or red) shifted emission bleeds into the red (or blue) shifted emission for both the H$_2$ and $^{12}$CO ALMA emission. This could be due to disk rotation (see Tyagi et al. in prep).


\begin{figure*}
\centering
    \includegraphics[width=0.9\linewidth]{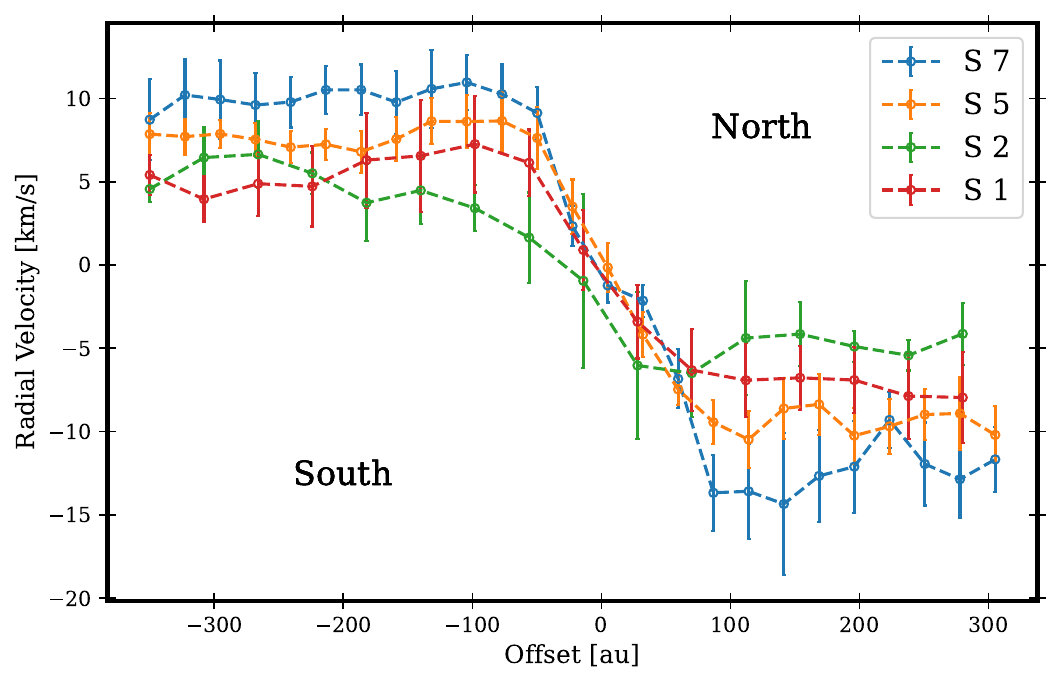}
\caption{Position velocity (PV) diagram for the H$_2$ 0-0~S(1), S(2), S(5) and S(7) lines extracted along the outflow direction. The error bars on the line are the errors in determining the velocity center of the line in each bin. We have set the mean flow velocity to zero due to uncertainties in the absolute velocity calibration. }
\label{Figure13}
\end{figure*}

Next, we constructed position velocity (PV) diagrams for the H$_2$ 0-0~S(1) to S(7) lines that were detected in MIRI MRS after subtracting the cloud velocity. As stated before, we found that there is an extended component to the H$_2$ emission (see Figure \ref{Figure4}). This component might be tracing the ambient emission from the cloud. This cloud component has a smaller velocity than the outflow component from the protostar. We used two 0.8\arcsec{} radius apertures on either side of the outflow (with offsets of $\delta_1$RA = -2.8\arcsec; $\delta_1$Dec =0.8 and $\delta_2$RA = 2.8\arcsec; $\delta_2$Dec =-0.8, with respect to 14 \micron{} position) to compute the velocity of this weak, slow cloud component. We fit a gaussian to this emission and derive the cloud emission, which we subtract from the data-cube pixel-by-pixel to remove the cloud component and derive the true velocity. After subtracting this ambient component, we extracted velocities from a rectangular region that was 5\arcsec{} long and 4\arcsec{} wide and aligned along the outflow direction (PA = 23\arcdeg) and centered at the 14\micron{} position. {To remove any velocity offsets arising from instrumental or systematic effects, we subtract the mean velocity of the PV diagram, which we define as the mean flow velocity. All velocities measured from the PV diagram are therefore expressed relative to this mean flow velocity.} In Figure \ref{Figure13}, the PV diagrams for the various molecular H$_2$ 0-0~S(J) lines are shown. 

The low excitation H$_2$ 0-0~S(1) and H$_2$ 0-0~S(2) lines have radial velocities in the range of $\sim$ $\pm$5.5~km~s$^{-1}$, while the higher transitions such as H$_2$ 0-0~S(5) and H$_2$ 0-0~S(7) have a radial velocity range between $\sim$~-10 to 7~km~s$^{-1}$ and  $\sim$~-15 to 10~km~s$^{-1}$ respectively.  This suggests that the higher excitation H$_2$ (hotter component of the outflow) is moving faster than the cooler low excitation gas component of the outflow as traced in low-$J$ H$_2$. If we further correct for inclination (with inclination angle of 64\arcdeg), we find that the low-$J$ H$_2$ outflow (S(1) and S(2)) has a velocity of $\sim$12.5~km~s$^{-1}$, while the hotter component as traced by  H$_2$ 0-0~S(5) and S(7) has a velocity of 28.5~km~s$^{-1}$. There could be a even faster component of the  H$_2$ outflow but due to the poorer velocity resolution of NIRSpec we are not able to detect it.

\begin{figure*}
\centering
\includegraphics[width=0.25\linewidth]{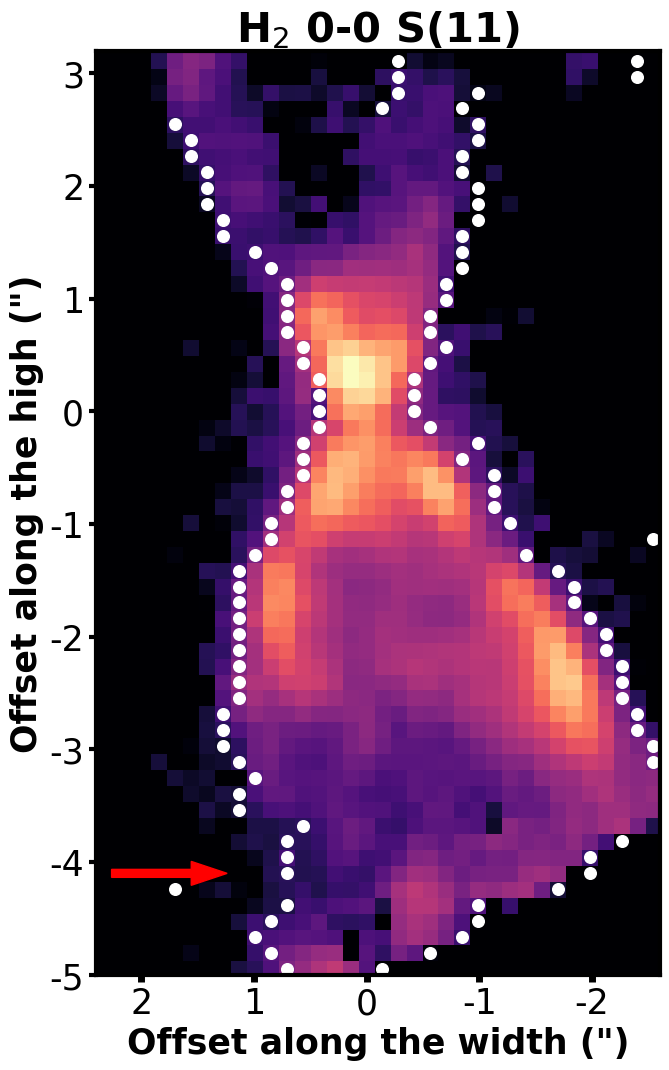}\includegraphics[width=0.25\linewidth]{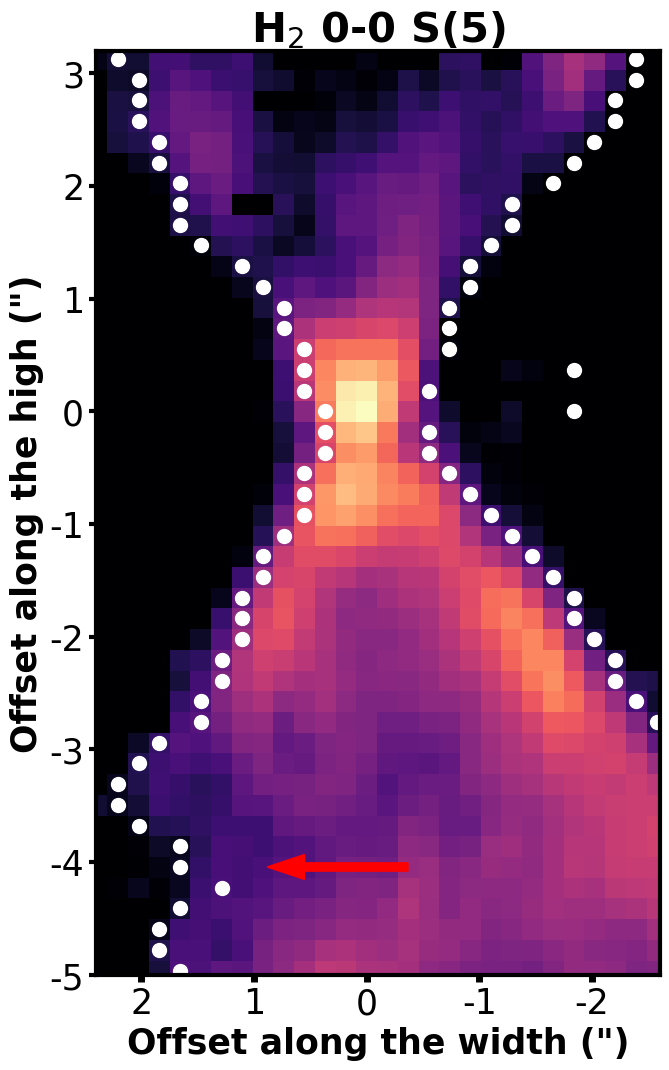}\includegraphics[width=0.25\linewidth]{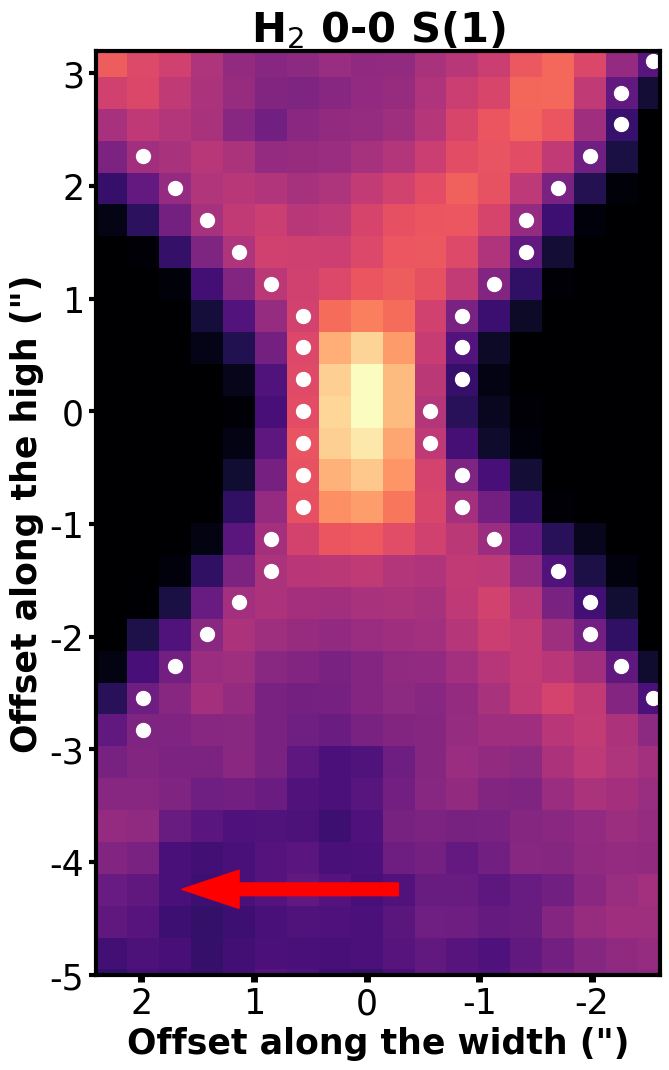}\includegraphics[width=0.25\linewidth]{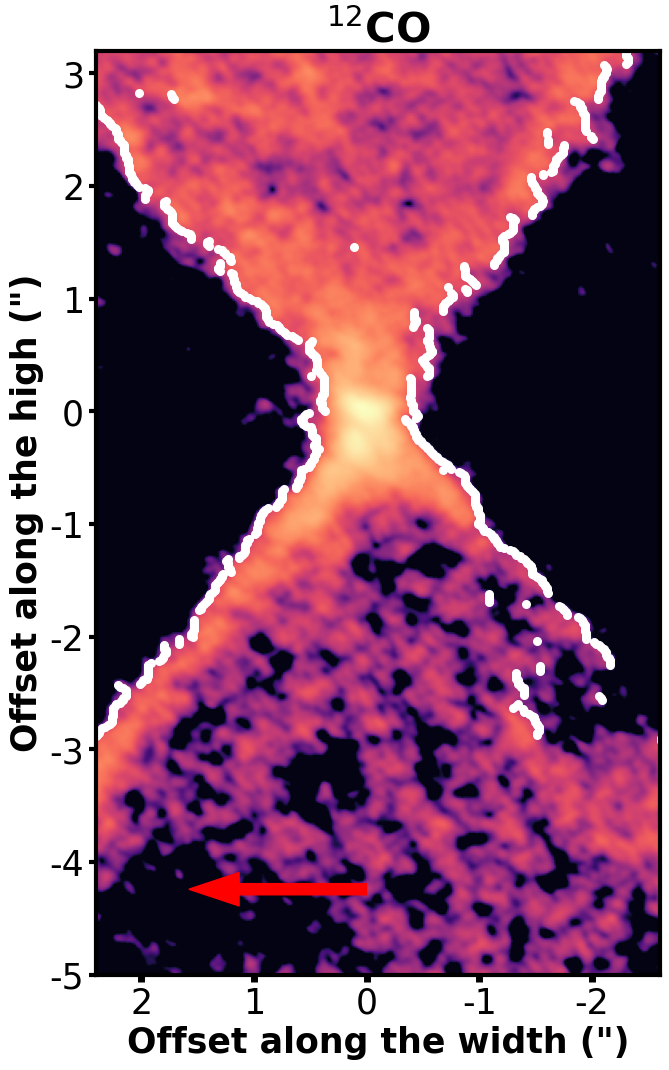}
\caption{An illustration of the detected edges of the molecular outflow. The color scale is the outflow and the edge is shown as white points. All images are cropped to the same area. We show a red arrow that illustrates the position of the notch in the outflow.  }
\label{Figure11}
\end{figure*}

\subsection{Opening angle of the outflow}

From Figure \ref{Figure6} and Figure \ref{Figure8}, it is evident that the different tracers of the outflow (with different excitation energy) have different spatial extents. However, to quantify this the difference in opening angle and the width of the outflow as a function of distance from the driving source need to be measured. To do so we first need to delineate the outer edge of the emission.  We adopted a method similar to that of \cite{2021ApJ...911..153H}, where we initially rotated the image by 23\arcdeg{} clockwise to align the outflow axis with the positive (y)-axis in an x - y Cartesian plane. The rotation angle, 23\arcdeg{}, was calculated from (PA of the disk) - 90\arcdeg{}, and the PA of the disk, 113\arcdeg{}, was determined by modeling the disk and is measured from the east of north  (\citealt{2023ApJ...954..101A} and Narang et al. in prep).  We computed a second-order differential along the x-axis, and the inflection points (where the second-order derivative equals zero) delineate the edge of the cavity. Using this approach, we obtained four cavity edges (two for each of the blue- and red-shifted outflows/cavities). A similar method was also used by \cite{2024arXiv241018033P} to measure the opening angle of the outflow for four edge-on sources. To avoid bad pixels and low S/N data we only used pixels with S/N $>7\sigma$.

In Figure \ref{Figure11}, we present the cavity edge as traced by this method. Our analysis reveals that the cavity edges deviate from a straight line and exhibit structural features. One prominent feature observed in the molecular H$_2$ 0-0 S(5) and S(11) emission is a notch in the outflow, located at $\sim$ -4\arcsec{} offset along the height of the outflow to the south. This same notch region corresponds to an emission minimum in both H$_2$ 0-0 S(1) and ALMA $^{12}$CO. In addition, H$_2$ 0-0 S(5) and S(1) show an outflow neck within $\sim$1\arcsec{} on both sides of the protostellar position, where the edge is almost parallel to the y-axis, and a similar neck is also seen in $^{12}$CO within $\sim$0.\arcsec5 on both sides.

To measure the opening angle of the outflow, we determined the half-width of the cavity as a function of distance from the central protostar (indicated as offset along the height in Figure \ref{Figure11}) by fitting a straight line. The inner 1\arcsec{} on both sides of the protostellar position was masked to exclude the outflow neck. {We further deconvolve the measurements by subtracting, in quadrature, the maximum beam/PSF radius from both the half-width of the cavity and the offset measured along the cavity height.}  In Table \ref{Table2} we have listed the {(deconvolved)} opening angle for the flow traced by H$_2$ 0-0 rotational lines with high S/N along with the ALMA $^{12}$CO. {We find a gradual decrease in the opening angle of the outflow with an increase in E$_{up}$. Specifically, the opening angle decreases from 40–35\arcdeg{} for the low-J H$_2$ lines (up to S(5)) and the cold gas component (ALMA $^{12}$CO) to $\sim$28–19\arcdeg{} for the high-J H$_2$ lines (S(7)–S(11)). }


\begin{deluxetable}{ccc}
\label{Table2}
\tablecaption{{The (deconvolved) opening angle, measured as the half-width of the outflow as a function of distance from the central protostar, derived for the high–S/N H$_2$ lines and the ALMA $^{12}$CO outflow.} }
\tablehead{\colhead{} & \multicolumn{2}{c}{Opening Angle (\arcdeg)} \\
\colhead{Line} & \colhead{Northern Outflow} & \colhead{Southern Outflow} } 

\startdata
$^{12}$CO J=2--1 & 38 $\pm$ 0.3 & 40  $\pm$ 0.3 \\
S(1) & $41 \pm 1$ & $40 \pm 1$ \\
S(2) & $38 \pm 1$ & $35 \pm 1$ \\
S(3) & $39 \pm 1$ & $33 \pm 1$ \\
S(4) & $39 \pm 2$ & $36 \pm 2$ \\
S(5) & $36 \pm 1$ & $34 \pm 1$ \\
S(7) & $25 \pm 4$ & $29 \pm 1$ \\
S(11) & $21 \pm 3$ & $19 \pm 2$ \\
\enddata

\end{deluxetable}

\subsection{Width of the Outflow at the the Protostar}

\begin{figure*}
\centering
\includegraphics[width=0.4\linewidth]{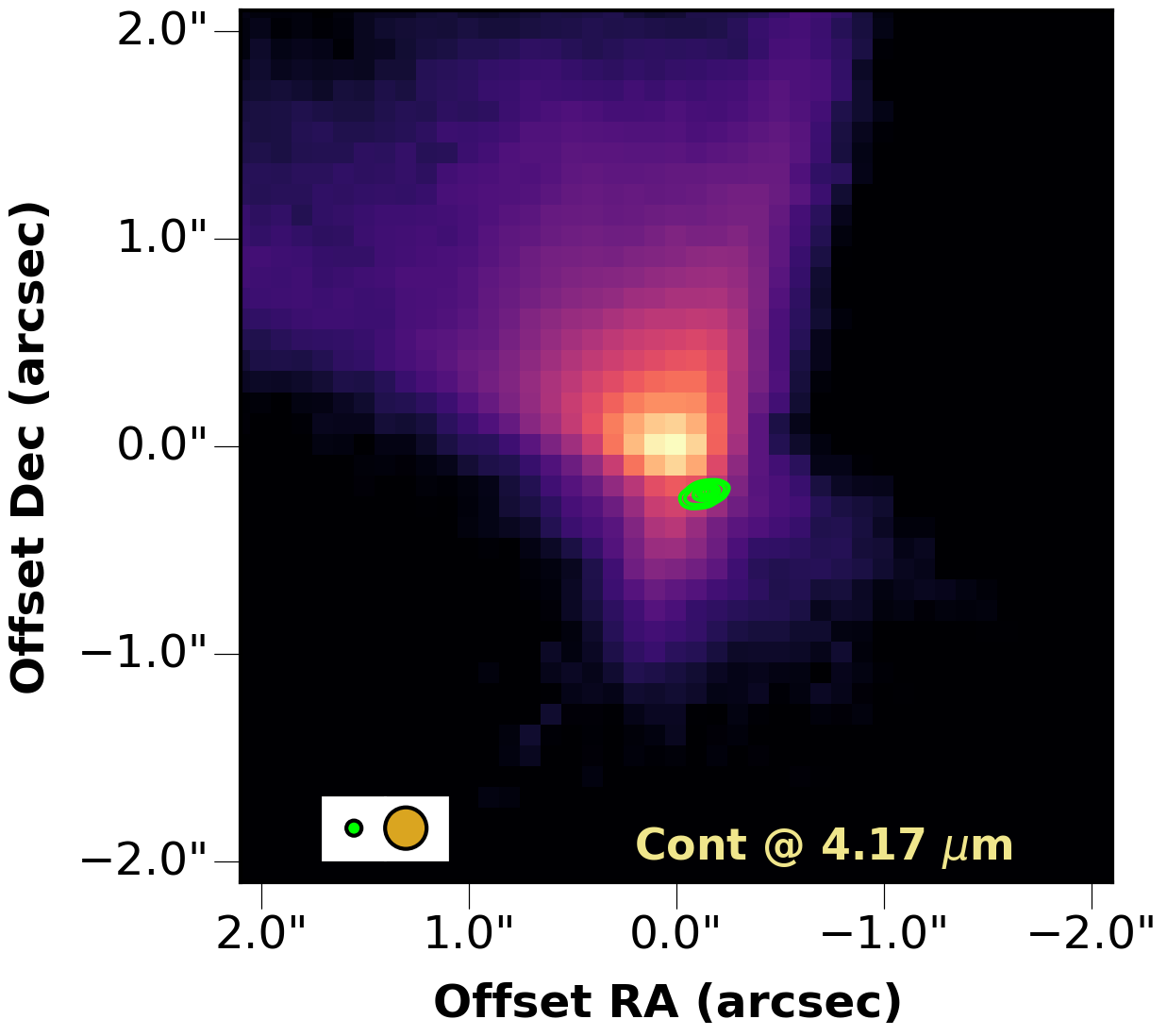}\includegraphics[width=0.4\linewidth]{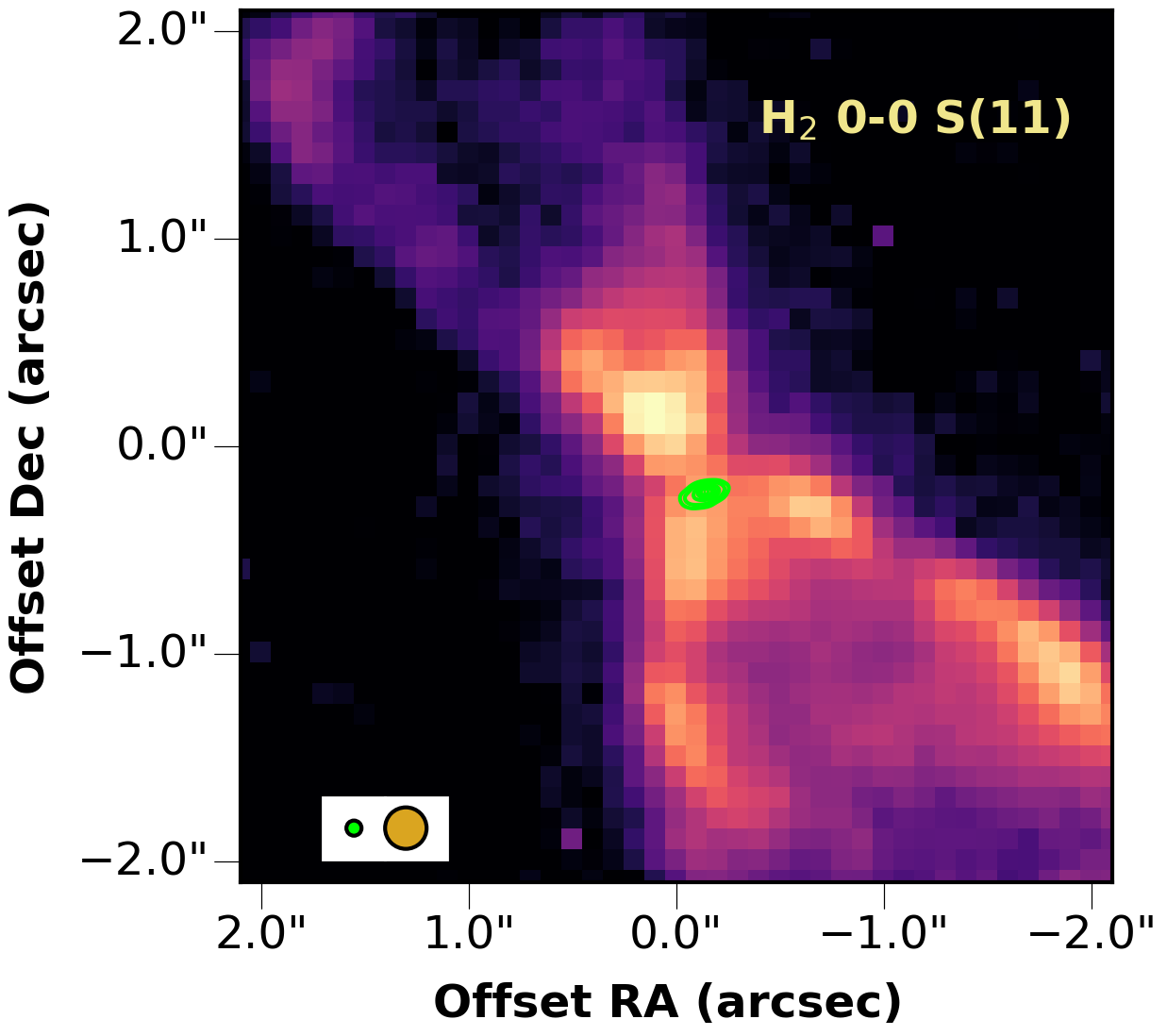}
\includegraphics[width=0.4\linewidth]{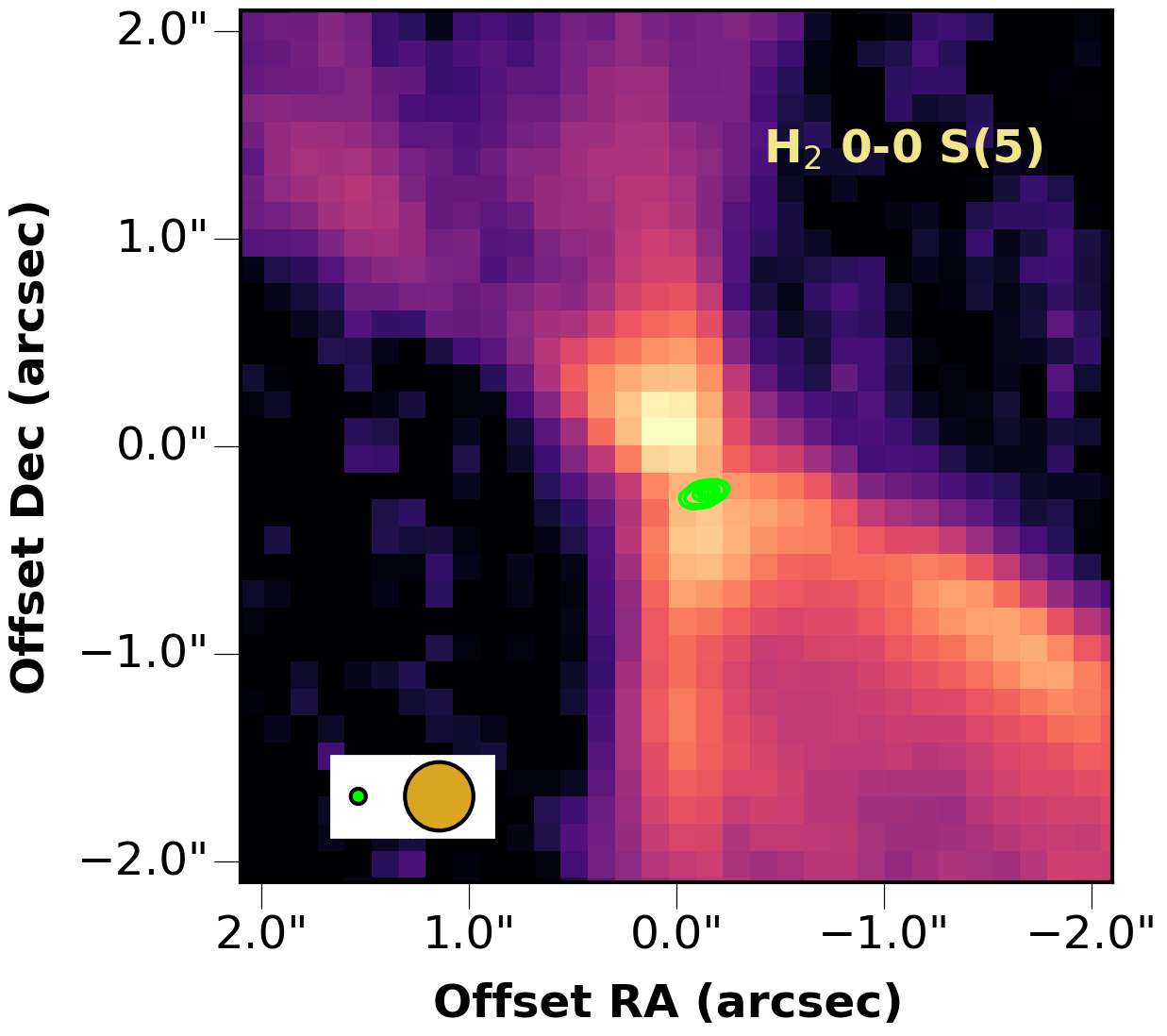}\includegraphics[width=0.4\linewidth]{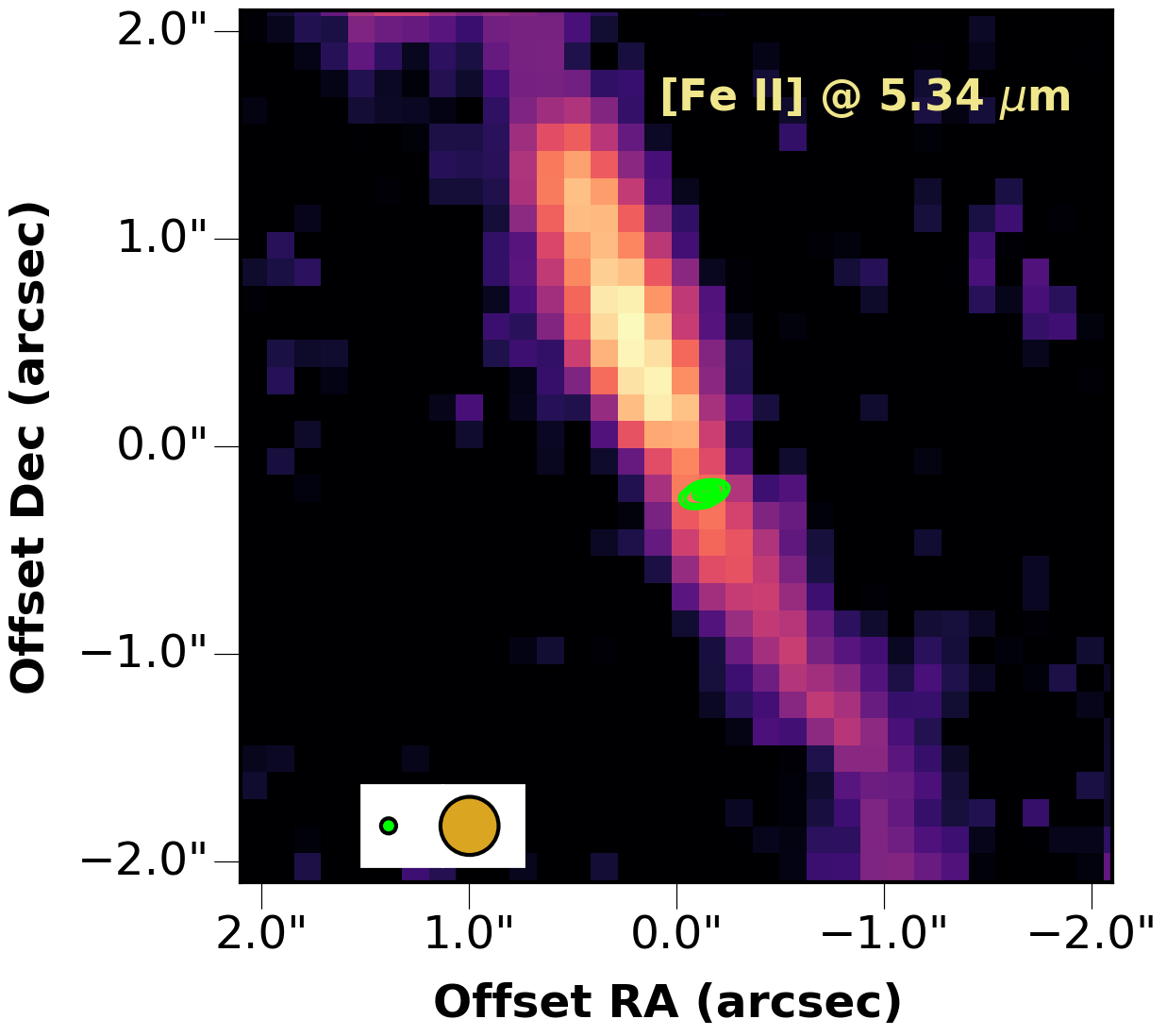}
\caption{The ALMA 1.3~mm continuum (lime contours) overlaid on top of the JWST continuum at 4.17~\micron{},  H$_2$ 0-0~S(11), H$_2$ 0-0~S(11) and the [Fe~II] line at 5.34~\micron{} in color scale.  The 1.3 mm contours are 10\%, 20\%, 40\%, 60\%, 80\% 99\% $\times$ 5.1~mJy/beam. {The JWST PSF as khaki circle and the maximum ALMA continuum beam  (of 0.073\arcsec) as lime circle are shown in the bottom left corner.}  }
\label{Figure9}
\end{figure*}

Figure \ref{Figure9} presents the ALMA 1.3~mm continuum contours superimposed on the JWST continuum map at 4.17~\micron{}, alongside the line maps for H$_2$ 0-0 S(5), H$_2$ 0-0 S(11), and [Fe II] at 5.34~\micron{} lines. We observe that the disk traced by the 1.3 mm dust continuum is not only narrower than the scattered light cavity, but also smaller than the H$_2$ outflow and the jet traced by [Fe II] at 5.34~\micron{}. Given that the ALMA observations of the continuum disk have a much higher resolution (0.\arcsec073~$\times$~0.\arcsec048) than the JWST observations, deconvolution of the MIRI and NIRSpec images is necessary to accurately measure the true width of the outflow at its base. 

To quantify the width of the H$_2$ outflow at the base of the protostar (14 \micron{} position) and compare it with the protostellar disk (both dust and gas), we adopt a method similar to \cite{2023arXiv231014061N}. We took slices perpendicular to the jet/outflow axis with a width of 2 pixels  (NIRSpec beam is 0.\arcsec2 or 2 pixels). Subsequently, each slice was fitted with a Gaussian, and the Full Width at Half Maximum (FWHM) of the Gaussian was measured as the width of the outflow in that slice. To measure the width of the outflow, we used the H$_2$ 0-0~S(11) line due to its high S/N and spatial resolution.

\begin{figure*}[ht]
\centering
\includegraphics[width=0.9\linewidth]{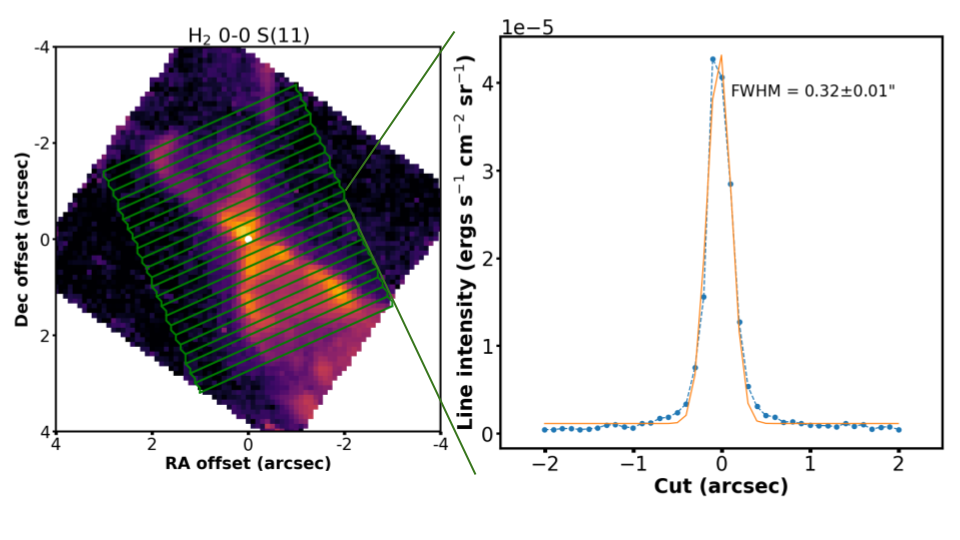}
\caption{(a) The slices aligned with the jet and overlaid in green on top of the molecular outflow traced in H$_2$ S(11) at 4.18~$\mu$m within which the width of the outflow is measured.  The white circle represents the 14 \micron{} protostellar position. (b) The Gaussian fit to the intensity profile for the slice at the protostellar position. }
\label{Figure10}
\end{figure*}

In Figure \ref{Figure10}(a), we show the H$_2$ 0-0~S(11) emission map with the slices marked in green.   We measure the width of the slice closest to the protostar and show it in  Figure \ref{Figure10}(b). The width of the outflow from the fit is $\sim$~0.32~$\pm$~0.01~\arcsec. This is larger than the PSF of NIRSpec IFU at 0.2\arcsec.  We deconvolve the outflow width using the NIRSpec IFU average PSF and obtain a deconvolved outflow width of $\sim 0.25$\arcsec. At the distance of IRAS~16253$-$2429 this translates into 35 au.  {However, this width should be regarded as an upper limit for several reasons. First, the measurement is averaged over two pixels, which inherently broadens the observed profile, even though we have subtracted out the PSF. Second, the outflow cross-section does not follow an exact Gaussian shape, so fitting it as a Gaussian can overestimate its true width. Together, these factors reinforce that the derived width represents a conservative upper bound rather than a precise physical measurement.}

The H$_2$ outflow width is much larger than the width of the [Fe~II] emission (at 5.34~\micron{})  of $20-23$~au \citep{2024ApJ...966...41F, 2023arXiv231014061N}.  {The ALMA 1.3~mm continuum dust disk and the Keplerian gas disk (tarced by $^{12}$CO) associated with IRAS~16253$-$2429 are about  30 au wide \citep{2023ApJ...954..101A}  which is slightly smaller than but comparable to the width of the outflow as traced by  H$_2$. }


\section{Origin of the outflow from IRAS 16253-2429}

Our observations of IRAS 16253$-$2429 using JWST and ALMA reveal a nested outflow morphology. The hourglass shape first detected by Spitzer IRAC is traced in gas  by the bipolar wide-angle outflows seen in H$_2$ and $^{12}$CO in the inner $\sim$ 2000 au of the protostar. The outflow cavity walls are best traced by the NIRSpec continuum and the cold $^{12}$CO emission from ALMA. Near the base of the protostar, the width of the molecular H$_2$ emission, as traced by the 0-0~S(11) line, is slightly smaller than the diameter of the continuum disk and the diameter of the Keplerian gas disk (traced in $^{12}$CO) around the protostar.

{When examining the flow traced by various H$_2$ transitions, we find that the higher-J transitions  (of S(7) and S(11) with E$_{up} >$ 5000 K) trace a much narrower outflow (based on the opening angle) as compared to the low-$J$ H$_2$ transitions  (E$_{up} \leq$ 5000 K) and ALMA $^{12}$CO J = 2-1 (E$_{up} \sim$ 30 K) (see Table \ref{Table2}). As the E$_{up}$ of the molecular tracer in the outflow increases, the outflow's opening angle narrows from $\sim$ 40-35\arcdeg{} for $^{12}$CO and low-J H$_2$ to 28-19\arcdeg{} for high-J H$_2$.}  This molecular outflow surrounds a highly collimated (opening angle of 2.\arcdeg6 $\pm$ 0.\arcdeg5) atomic/ionic jet detected in multiple [Fe II] transitions, as well as [Ne II], [Ni II], [Ar~II]  and H-I (Br$\alpha$) with JWST \citep{2023arXiv231014061N, 2024ApJ...966...41F}.

Similar outflow and jet morphology have been observed in the Class I protostar TMC 1E \citep{2024A&A...687A..36T},  where the fine-structure jet is highly collimated and is surrounded by a much wider molecular outflow. \cite{2024A&A...687A..36T} found that the degree of collimation increases with increasing E$_{up}$ for the molecular H$_2$ and that the molecular H$_2$ outflow was more collimated than the ALMA $^{12}$CO outflow and scattered light cavity, similar to what we find for IRAS 16253-2429.  Such a nested morphology was also detected in  DG Tau B (also a Class I protostar) \citep{2024arXiv240319400D} using a combination of ALMA, JWST, and ground-based VLT observations with SINFONI. 

{Recently, \cite{2024arXiv241018033P} reported similar results for the edge-on source HH30, finding a highly collimated jet surrounded by molecular H$_2$ and a cold $^{12}$CO outflow. They observed that the H$_2$ emission was more collimated than the $^{12}$CO outflow detected with ALMA. Based on the combination of this nested morphology and a small launch radius ($\sim$7 au), \cite{2024arXiv241018033P} argued that radially extended magnetohydrodynamic (MHD) disk winds were the only mechanism consistent with the nested jet and outflow structures observed in their sample.}


In contrast to the more evolved protostars and Class II disks in the previous studies, IRAS 16253-2429 is in the envelope dominated, Class 0 stage, where most of the mass accretion occurs \citep{2017ApJ...840...69F}. It is in this phase where the interaction of disk launched winds and jets with the  infalling envelope entrains gas and regulates the masses of stars \citep{2023ApJ...947...25H}.  This interaction must occur along the cavity walls. Our data on the Class 0 protostar IRAS 16253-2429 clearly show a wide angle wind traced by the H$_2$ is directly interacting with the envelope along the cavity walls.  We see the H$_2$ emission extend through the cavities.  Along the cavity walls, we see enhanced H$_2$ emission in the lower rotational lines.  In addition, in certain areas where the shocks may be less oblique, along a notch in the northern cavity and a curved portion of the southern cavity, we see shocks along the cavity walls in higher transitions. The less oblique shocks in these regions appear to result in warmer shocked gas.  Given the lower velocity of the ALMA CO emission \citep{2023ApJ...954..101A}, this component likely traces entrained gas accelerated along the cavity walls, although further analysis is required to rule out a contribution from warm gas in a disk wind.

The MHD disk wind naturally explains the presence of a primarily molecular component of the flow \citep{2012A&A...538A...2P}.  Evidence for disk winds being the driving mechanism behind the nested structure of IRAS 16253-2429 comes from the analysis of the kinematics of the molecular outflow as traced in H$_2$. The position-velocity diagrams reveal a clear distinction in the kinematics of the low- and high-excitation H$_2$ transitions. The low-$J$ H$_2$ transitions exhibit relatively low velocities, $\sim$ 12.5 km s$^{-1}$ with respect to the mean flow velocity. In contrast, the higher-$J$ transitions, such as H$_2$ 0-0~S(5) and S(7), display significantly higher velocities, around 28.5 km s$^{-1}$ with respect to the mean flow velocity. This suggests that the higher-energy H$_2$ emission, not only appears more collimated but is also moving considerably faster compared to the wider, slower-moving low-$J$ H$_2$ emission. This is consistent with the results of MHD wind simulations, which predicts faster speeds for more collimated parts of the flow (e.g.,\citealt{1983ApJ...274..677P,1986ApJ...301..571P,1993ApJ...410..218W,2000prpl.conf..759K,2007prpl.conf..277P,2023ASPC..534..567P}).   
 
In contrast, X-wind models predict that both the jet and the wide-angle wind components are launched from the innermost regions of the disk and therefore should exhibit similar velocities. Within the X-wind framework \citep{1994ApJ...429..808N,2007prpl.conf..261S}, the slower H$_2$ emission we observe would be tracing shocked material within the wind. To reproduce the observed flow, X-wind models must account for both the molecular nature of the warm, slower component and the atomic or ionic composition of the fast, collimated jet, which some recent simulations  have begun to achieve \citep[e.g.,][]{2020ApJ...905..116S,2023ApJ...945L...1S} {by introducing the interaction of X-winds with the ambient material.}

{Recently, \citet{2025arXiv251100515N} demonstrated that photoevaporative winds can exhibit morphological characteristics similar to those of MHD winds, with higher-$J$ H$_2$ transitions displaying a more collimated structure. However, MHD disk–wind models predict predominantly molecular outflows \citep[e.g.,][]{2012A&A...538A...2P}, whereas the strong high-energy radiation required to drive photoevaporative winds can efficiently dissociate H$_2$, resulting in very low molecular fractions \citep{2017ApJ...847...11W}. IRAS~16253-2429 is currently in a quiescent phase and is therefore unlikely to possess a sufficiently strong UV field to drive photoevaporation. Moreover, the models of \citet{2025arXiv251100515N} predict that photoevaporative winds typically reach velocities of $\lesssim 10~\mathrm{km,s^{-1}}$, whereas for IRAS~16253-2429 we measure significantly higher velocities, up to $\sim 20~\mathrm{km,s^{-1}}$, albeit with large uncertainties. Finally, the presence of highly collimated atomic/ionic jets traced by e.g., [Fe II], [Ne II], and [Ar II]  \citep{2023arXiv231014061N} further supports an MHD wind scenario over a photoevaporative origin.}

Furthermore, some of the substructures observed in the higher transitions  of H$_2$ such as the one associated with the jet knot (see Figure \ref{Figure4}) can been interpreted as a bow-shock and wake driven by internal shocks in the jet \citep{2023arXiv231014061N,2024ApJ...966...41F}. These structures can be generated when shocks in jets with fluctuating velocities result in material ejected laterally from the jet \citep{2018A&A...614A.119T, 2022A&A...664A.118R}.  These require a slow moving wind surrounding the jet \citep{2018A&A...614A.119T}, and they can accelerate the material in the wind \citep{2021ApJ...907L..41L}. Thus, the high velocity, more collimated component may be further accelerated  by interactions with the jet \citep{2024arXiv240319400D, 2022A&A...664A.118R}.  We further speculate that the shear in the wind flow, both from the interaction with the jet and due to the range of launching speeds in the nested structure, may be responsible for heating of the gas by shocks.

\section{Summary}

We report the results from joint JWST and ALMA observations of the Class~0 protostar IRAS~16253$-$2429 in the Ophiuchus molecular cloud (140~pc) that was observed as part of the ALMA large program ``Early Planet Formation in Embedded Disks (eDisk)" and the JWST Cycle~1 GO project ``Investigating Protostellar Accretion (IPA)". The key findings from our study are summarized as follows:

\begin{enumerate}
    \item  The continuum cavity traced in the scattered light exhibits a wide hourglass-like morphology, with the northern side appearing brighter and extending further compared to the southern side due to extinction and inclination effects.  This cavity follows the shape of the entrained outflow detected with ALMA except at the neck of the hourglass, where the scattered light is wider.  At longer wavelengths where scattering is less efficient, the IR emission peak is concentrated toward the central source, but with a small offset between the emission peak and the ALMA disk position.   
    
    \item Within the cavity we detect 15 lines of pure rotational molecular H$_2$ (v=0-0) emission ranging from S(1) to  S(15), along with a few  H$_2$ 1-0~O(J) and  H$_2$ 1-1~S(J) transitions. This emission fills the cavity in the lower transitions, indicating that the cavity is filled with warm, molecular gas. 

    \item The position velocity diagrams show that low E$_{up}$ H$_2$ lines (0-0~S(1) and S(2)) have an inclination corrected velocity of $\sim$12.5~km~s$^{-1}$ while higher E$_{up}$ lines such as H$_2$ 0-0~S(5) and S(7) have an inclination corrected velocity of $\sim$28.5~km~s$^{-1}$. This velocity structure shows the warm molecular gas filling the cavity is tracing a moderate velocity outflow, slower than the central jet seen in atomic / ionic species.

    \item {A clear trend is observed in which the outflow opening angle decreases with increasing E$_{up}$. The angle reduces from 40–35\arcdeg{} for the low-J H$_2$ transitions (up to S(5)) and the cold gas traced by ALMA $^{12}$CO, to $\sim$28–19\arcdeg{} for the higher-excitation H$_2$ lines (S(7)–S(11)).}

    \item {The outflow (detected in H$_2$ 0-0~S(11)) is $\leq$35 au wide at the protostellar location. This is slightly larger than but comparable to the diameter of the dust disk (30~au)  at 1.3~mm as detected with ALMA as well as the Keplerian disk diameter of $\sim$ 32~au. The width of the outflow however is much wider than the jet width of 20--23~au (as seen in the [Fe~II] lines)}.  

    \item The observed nested morphology and the kinematics of the jet (as traced in [Fe II] and atomic/ionic lines) and the extended molecular outflow traced in H$_2$ suggest MHD disk-winds as a possible launching mechanism. The faster, narrower emission in the higher excitation lines appears to be tracing a wake generated by internal shocks within the jet. Thus some of the acceleration of the inner wind may be due to processes in the jet. 

    \item The $^{12}$CO observed by ALMA is slower and colder than the molecular wind traced in the warm H$_2$ gas. This gas may be, at least in part, the entrained gas along the cavity walls accelerated by the molecular wind.  Enhancements in low lying H$_2$ along the cavity edges likely trace oblique shocks tracing the interaction of the wind and entrained gas.  In two regions, along the curved wall in the southern cavity and a notch in the northern cavity, the detection of S(11) emission suggests the presence of hotter, more direct shocks.
    
\end{enumerate}

This study underscores the synergy between JWST and ALMA in investigating jets and outflows from protostars, demonstrating how these observations can help constrain the mechanisms behind these phenomena.

\section{Data availability}
All of the JWST data presented in this article were obtained from the Mikulski Archive for Space Telescopes (MAST) at the Space Telescope Science Institute. The specific observations analyzed can be accessed via \dataset[DOI: 10.17909/3kky-t040]{https://doi.org/10.17909/3kky-t040}. The ALMA observations were obtained as part of the ALMA large program  Early Planet Formation in Embedded Disks eDisk (2019.1.00261.L, 2019.A.00034.S). 

\section{Acknowledgment}
This work is based on observations made with the NASA/ESA/CSA James Webb Space Telescope. The data were obtained from the Mikulski Archive for Space Telescopes at the Space Telescope Science Institute, which is operated by the Association of Universities for Research in Astronomy, Inc., under NASA contract NAS 5-03127 for JWST. These observations are associated with program \#1802. This paper makes use of the following ALMA data: ADS/ JAO.ALMA\#2019.1.00261.L. ALMA is a partnership of ESO (representing its member states), NSF (USA), and NINS (Japan), together with NRC (Canada), MOST and ASIAA (Taiwan), and KASI (Republic of Korea), in cooperation with the Republic of Chile. The Joint ALMA Observatory is operated by ESO, AUI/NRAO, and NAOJ. 
The National Radio Astronomy Observatory is a facility of the National Science Foundation operated under cooperative agreement by Associated Universities, Inc. N.O. and M.N. acknowledge support from National Science and Technology Council (NSTC) in Taiwan (NSTC 113-2112-M-001-037) and Academia Sinica Investigator Project Grant (AS-IV-114-M02). Part of this research was carried out at the Jet Propulsion Laboratory, California Institute of Technology, under a contract with the National Aeronautics and Space Administration (80NM0018D0004). H.T. and P.M. acknowledge the support of the Department of Atomic Energy, Government of India, under Project Identification No. RTI 4002. Support for SF, AER, STM, RG,  JG, JJT, and DW in program \#1802 was provided by NASA through a grant from the Space Telescope Science Institute, which is operated by the Association of Universities for Research in Astronomy, Inc., under NASA contract NAS 5-03127. LWL acknowledges support from NSF AST-2108794. G.A. and M.O. acknowledge financial support from grants PID2020-114461GB-I00, PID2023-146295NB-I00, and CEX2021-001131-S, funded by MCIN/AEI/10.13039/501100011033. A.C.G. acknowledges support from: PRIN-MUR 2022 20228JPA3A “The path to star and planet formation in the JWST era (PATH)” funded by NextGeneration EU, INAF-GoG 2022 “NIR-dark Accretion Outbursts in Massive Young stellar objects (NAOMY)”, and Large Grant INAF 2022 “YSOs Outflows, Disks and Accretion: towards a global framework for the evolution of planet forming systems (YODA)”. Y.-L.Y. acknowledges support from Grant-in-Aid from the Ministry of Education, Culture, Sports, Science, and Technology of Japan (20H05845, 20H05844, 22K20389), and a pioneering project in RIKEN (Evolution of Matter in the Universe). RK acknowledges financial support via the Heisenberg Research Grant funded by the Deutsche Forschungsgemeinschaft (DFG, German Research Foundation) under grant no.~KU 2849/9, project no.~445783058. NJE thanks the Astronomy Department of the University of Texas for research support.  S.J.W. acknowledges support from the Smithsonian Institution and the Chandra X-ray Center through NASA contract NAS8-03060. 

\facility{JWST (NIRSpec, MIRI), ALMA}

© 2025. All rights reserved

\bibliography{JWST}

\end{document}